\documentclass[preprint, times]{elsarticle}
\usepackage{graphicx}
\usepackage[margin=1in]{geometry}
\usepackage{amsmath}
\usepackage{bm}
\usepackage{setspace}
\usepackage[compact]{titlesec}
\usepackage{xr-hyper}
\usepackage{hyperref}
\usepackage{float}
\usepackage[usenames,dvipsnames]{xcolor}

\newcommand{\abbreviations}[1]{%
  \nonumnote{\textit{Abbreviations:\enspace}#1}}

\begin{document}

\begin{frontmatter}
	\journal{arXiv}
\title{Corner accuracy in direct ink writing with support material}

\author[]{Leanne Friedrich\fnref{fn1}}
\author[]{Matthew Begley\corref{cor1}\fnref{fn1}}

\address[fn1]{Materials Department, University of California Santa Barbara, Santa Barbara, CA, 93106, USA.}
\cortext[cor1]{Tel: (805)-679-1122; E-mail: begley@engr.ucsb.edu}
\begin{abstract}
    3D printing methods which enable control over the position and orientation of embedded particles have promising applications in cell patterning and composite scaffolds. Extrusion-based additive manufacturing techniques such as fused deposition modeling and direct ink writing can experience particle patterning defects at corners which could hinder cell survival at corners and create unintended property gradients. Here, we propose models which predict the behavior of deposited lines at corners for moderate viscosity inks which are impacted by both capillarity and viscous dissipation. Using direct ink writing with acoustophoresis and a Carbopol-based support gel, we write polygons out of dental resin-based composite inks containing a narrow distribution of microparticles at the center of the filament. A Laplace pressure differential between the inner and outer surfaces of the corner drives corner smoothing, wherein the inner radius of the corner increases. Double deposition, or printing on the same area twice, drives corner swelling, wherein excess ink is diverted to the outer edge of the corner. Fast turns at corners produce ringing, wherein vibrations in the stage manifest in oscillations in the print path. Swelling and ringing effects are apparent in the particle distributions at corners immediately after deposition, while smoothing effects are apparent after the printed structure has had time to relax. When the nozzle returns to write a neighboring line, it imposes shear stresses which mitigate inconsistencies in microstructure at corners by erasing defects which appeared during relaxation. Using a support bath instead of layer-by-layer support suppresses microstructural corner defects.
\end{abstract}

\begin{keyword}
	Acoustic Focusing \sep Acoustophoresis \sep Corner smoothing \sep Corner swell \sep Direct ink writing \sep Polymer matrix composite
\end{keyword}

\abbreviations{
	DIW direct ink writing; TEGDMA triethylene glycol dimethacrylate; UDMA diurethane dimethacrylate
}

\end{frontmatter}

\section{Introduction}

3D printing can enable cell patterning and functionally graded scaffolds through field-assisted manipulation of multiphase inks. In addition to biological applications, extrusion-based additive manufacturing enables fabrication of structures from a wide range of material systems. Fused deposition modeling can be used to print viscous thermoplastics, while direct ink writing (DIW) can be used to print a diverse array of materials from low-viscosity inviscid biocompatible hydrogels\cite{Jin2017} to viscous nanoclay-epoxy composites\cite{Compton2014} to colloidal ceramic gels\cite{Lewis2006}. However, extrusion-based techniques inherently face challenges at corners of printed geometries. The shape and microstructure of the printed line change wherever the moving nozzle or moving stage changes direction. These changes at corners could degrade the functionality of printed composite structures. For example, when printing filaments with aligned and positioned cells,\cite{Armstrong2018, Sriphutkiat2019} distortion at corners could hinder cell adhesion and tissue formation. Alternatively, when printing self-insulated electrically conductive filaments,\cite{Melchert2019} defects in the printed microstructure at corners could create shorts in the conductive pathways. 

\begin{figure*}
	\centering\includegraphics[]{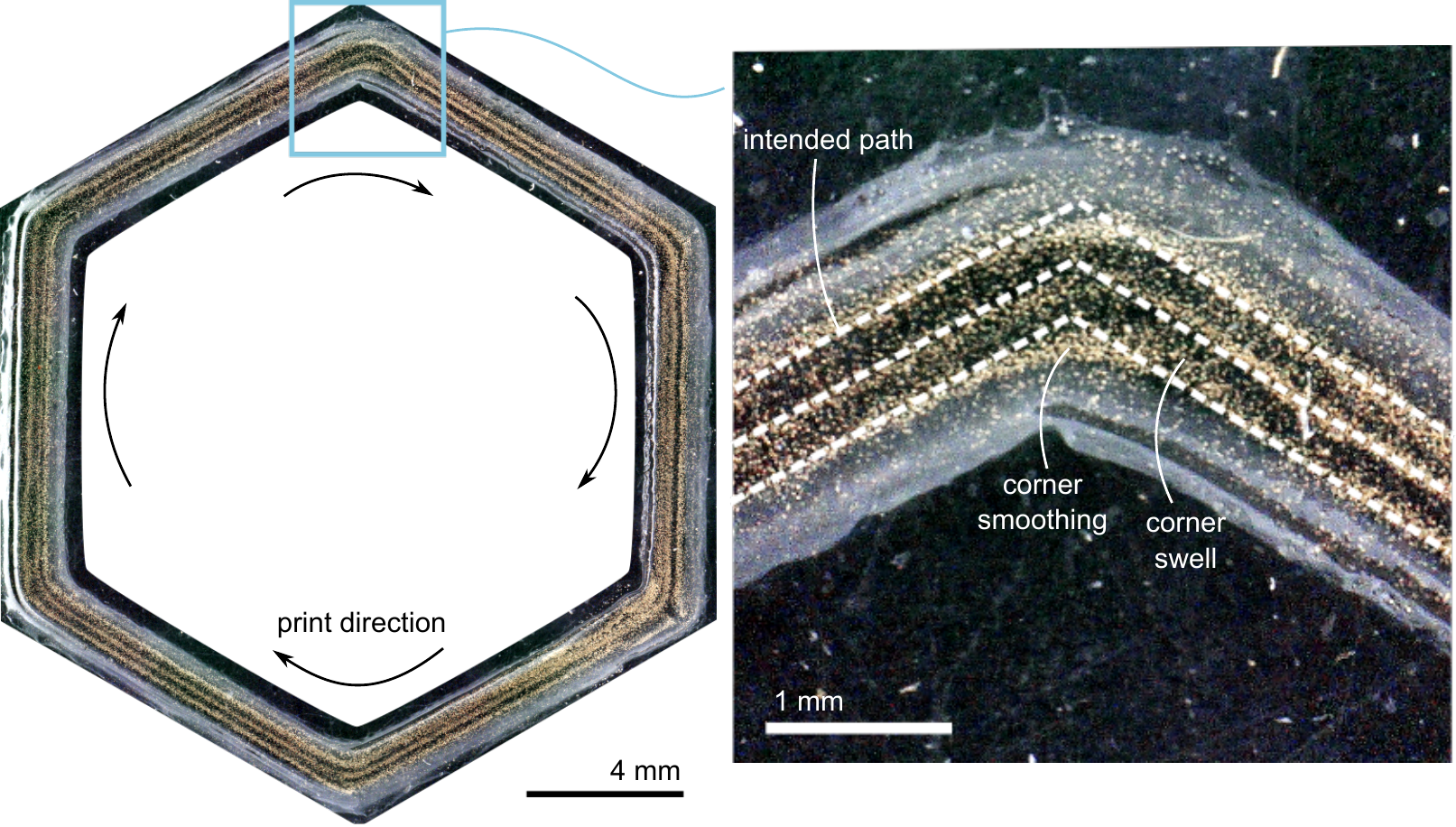}
	\caption[Printed corner]{A single layer hexagon printed with DIW with acoustophoresis experiences inaccurate particle positioning at corners due to corner smoothing.}
	\label{fig:prints}
\end{figure*}

Changes in microstructure at corners can be broken into three effects: smoothing, swelling, and ringing. Smoothing occurs when the printed line bends inward at the corner (Fig. \ref{fig:prints}), either because the printer has been programmed to traverse a blunt path to avoid swelling and ringing\cite{Comminal2019} or because the deposited fluid changes shape at the corner after deposition.\cite{Huang2017} Swelling occurs (Fig. \ref{fig:prints}) when excess fluid is generated at the corner because the nozzle must retrace its path at the corner, a pattern known as double deposition. Swelling can also occur when the translation speed is decreased at corners to avoid smoothing and ringing.\cite{Comminal2019, Kulkarni1999, Han2002} Ringing occurs because rapid changes in print direction can induce vibrations in the positioning gantry which manifest in ripples on the surface of the printed structure.\cite{Okwudire2018, Galati2019} Slower translation speeds limit the impacts of smoothing, swelling and ringing but also limit the throughput of extrusion-based techniques, so alternate strategies for limiting these effects are needed. 

Previously, strategies for mitigating the adverse effects of smoothing, swelling, and ringing have been studied for fused deposition modeling, where inks are viscous enough that capillarity can be ignored.\cite{Comminal2019, Okwudire2018, Han2002, Galati2019} Strategies for mitigating the adverse effects of smoothing have been studied for direct liquid writing, where ink viscosities are low enough that viscous dissipation can be ignored.\cite{Huang2017} However, DIW encompasses inks with a wide range of rheological properties, and a model is needed to describe the corner behavior of inks within a viscosity range where viscous dissipation and interfacial energies are both significant, such as the dental resin-based inks used herein or many hydrogels. Here, we describe a three-part model that describes changes in the printed line at corners as a result of interfacial energy-driven smoothing, double deposition-driven swelling, and ringing due to rapid acceleration.

\begin{figure*}
	\centering\includegraphics[]{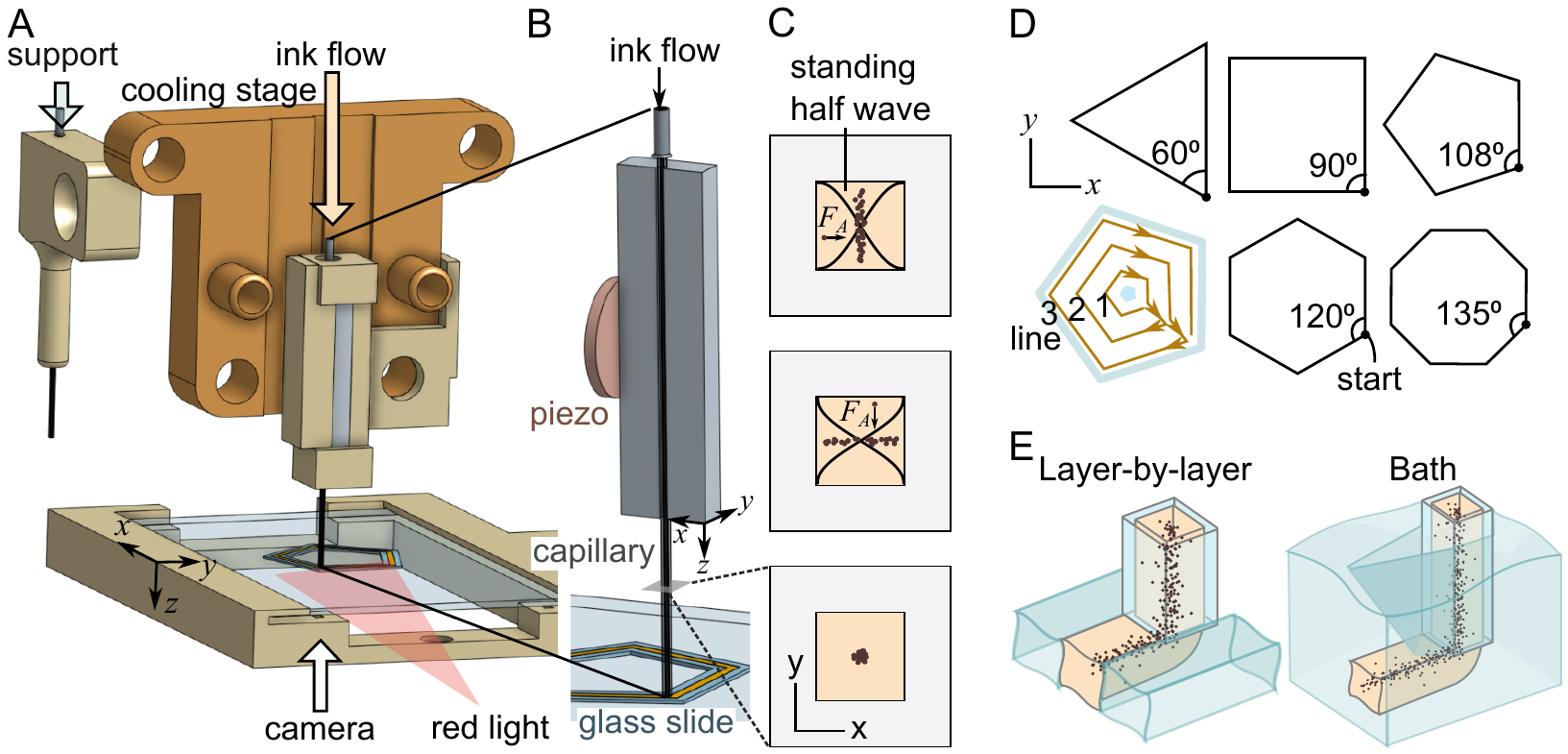}
	\caption[Schematic of printer, polygons, and support]{A) Using layer-by-layer support, the stage travels between support and ink nozzles while videos are collected from underneath the nozzle. $x$, $y$, and $z$ refer to G-code coordinates. B) Ink is extruded onto a glass slide through a glass capillary seated in a stainless steel block glued to a piezo. C) Two standing half-waves in the $x$ and $y$ directions drive particles with acoustic force $F_A$ toward the center of the channel. D) Printed polygons probe five corner angles. E) Layer-by-layer support is deposited in a specific pattern, while bath support fills the entire print envelope.}
	\label{fig:printdiag}
\end{figure*}

To characterize the corner behavior of moderate-viscosity direct-write inks, we used DIW with acoustophoresis to write dental resin-based composite lines containing a narrow distribution of metallic microparticles at the center. These materials serve as a model system that is generally representative of high density particles in a low density matrix. In DIW with acoustophoresis, a piezoelectric transducer attached to the print nozzle establishes a standing bulk acoustic wave inside of the nozzle. Given a sufficient acoustic contrast factor (which depends on the densities and compressibilities of the particles and fluid matrix), the particles align and move toward the nodes or antinodes of the standing wave.\cite{Laurell2007, Collino2015a, Collino2016, Collino2018} Acoustophoresis has been used to align and position particles ranging from cells to carbon fibers to metallic microspheres.\cite{Melchert2019, Armstrong2018, Collino2016} Here, we use a square glass capillary as a print nozzle and dense metallic microparticles in a polyurethane-based matrix, so standing waves are generated in both directions transverse to the direction of flow, and particles move to a point in the center of the nozzle (Fig. \ref{fig:printdiag}C). The capillary is seated in a stainless steel channel holder glued to a piezoelectric transducer (Fig. \ref{fig:printdiag}B). The piezo is thermally coupled to a liquid-cooled stage which keeps the system at room temperature.\cite{Friedrich2017} The nozzle extrudes continuous lines of ink onto a glass slide mounted on a stage which moves in three dimensions (Fig. \ref{fig:printdiag}A). In doing so, we can write composite lines which contain a narrow distribution of microparticles at the center. We use the position and width of this particle distribution as a metric to track the shape of the line at the corner and corresponding changes in the microstructure of the line at the corner. 

The moderate-viscosity inks used in this study are shear thinning, but they do not exhibit solid-like behaviors at low shear stresses, which are critical for printing self-supporting structures.\cite{Lewis2006, Duty2018} To support the printed structures, we use a water-based Carbopol support gel which has been used as a support bath for similar low-viscosity inks including hydrogels, silicone elastomers, and liquid metal.\cite{Bhattacharjee2015, Jin2016, Leblanc2016, OBryan2018, Hinton2016} Carbopol gels act as Bingham plastics, so they exhibit elastic solid-like behaviors at low shear stresses and shear thinning liquid-like behaviors above a yield stress. The gels can be used as a support bath, where a submerged nozzle plastically deforms the bath locally, and the solid-like bath holds the printed line in place (Fig. \ref{fig:printdiag}E).\cite{Bhattacharjee2015, Jin2016, Leblanc2016, OBryan2018, Hinton2016} The gels can also be extruded layer-by-layer from a separate support nozzle to write designed shapes before ink deposition (Fig. \ref{fig:printdiag}E). Because the support geometry imposes different stresses on the printed ink, the two geometries produce different corner behaviors. In this work, we test both support geometries and find that bath support limits changes in particle distribution at the corner, improving microstructural uniformity across the print.

To test the influence of the sharpness of the corner on corner behaviors, we printed single-layer, three-pass equilateral triangles, squares, pentagons, hexagons, and octagons (Fig. \ref{fig:printdiag}D). This work focuses on how the particle distributions at the corners in the first pass change over the course of the print. We measure the difference in particle distribution position and width between the center of the polygon edge and the corner just after deposition, after the structure has had time to relax, and after the nozzle returns to write the second line and shears the first line. We find that corner swelling and ringing strongly influence the microstructure of the just-deposited line, and interfacial energy-driven corner smoothing changes the microstructure of the line during relaxation. Further, shear from the nozzle during deposition of subsequent lines can help to mitigate the effects of swelling, ringing, and smoothing in existing lines.

This work is part of a group of three papers which consider different aspects of the same data set. Ref. \cite{Friedrich2019s} considers printing direction-independent effects which manifest in straight lines. Ref. \cite{Friedrich2019d} considers direction-dependent effects in straight lines. This paper considers changes in microstructure at printed corners.

\section{Theory}

In this work, we probe the impact of corner angle $\theta$, print speed $v_s$, and ink composition on corner defects. In these experiments, the print speed represents both the flow rate in the nozzle and the translation speed of the stage, which are matched, although only the translation speed matters for these derivations. Herein, models are developed to predict two metrics: the differences in particle distribution peak position and width between the corner and center of the polygon edge:

\begin{equation}
    \Delta Position(\theta, v_s, ink) = Position(corner) - Position(center)
    \label{eq:dpos1}
\end{equation}

\begin{equation}
    \Delta Width(\theta, v_s, ink) = Width(corner) - Width(center)
    \label{eq:dwidth1}
\end{equation}

Both metrics are corner defects. $\Delta Position$ represents a path shift at the corner, while $\Delta Width$ represents spreading at the corner. With the following theories, we predict $\Delta Position(\theta, v_s, ink)$ and $\Delta Width(\theta, v_s, ink)$ due to smoothing, swelling, and ringing. Experimentally, we measure $\Delta Position(\theta, v_s, ink)$ and $\Delta Width(\theta, v_s, ink)$ and compare those values to the predicted values to determine the contributions of smoothing, swelling, and ringing.

\subsection{Interfacial energy-driven smoothing}

\begin{figure*}
    \centering
    \includegraphics{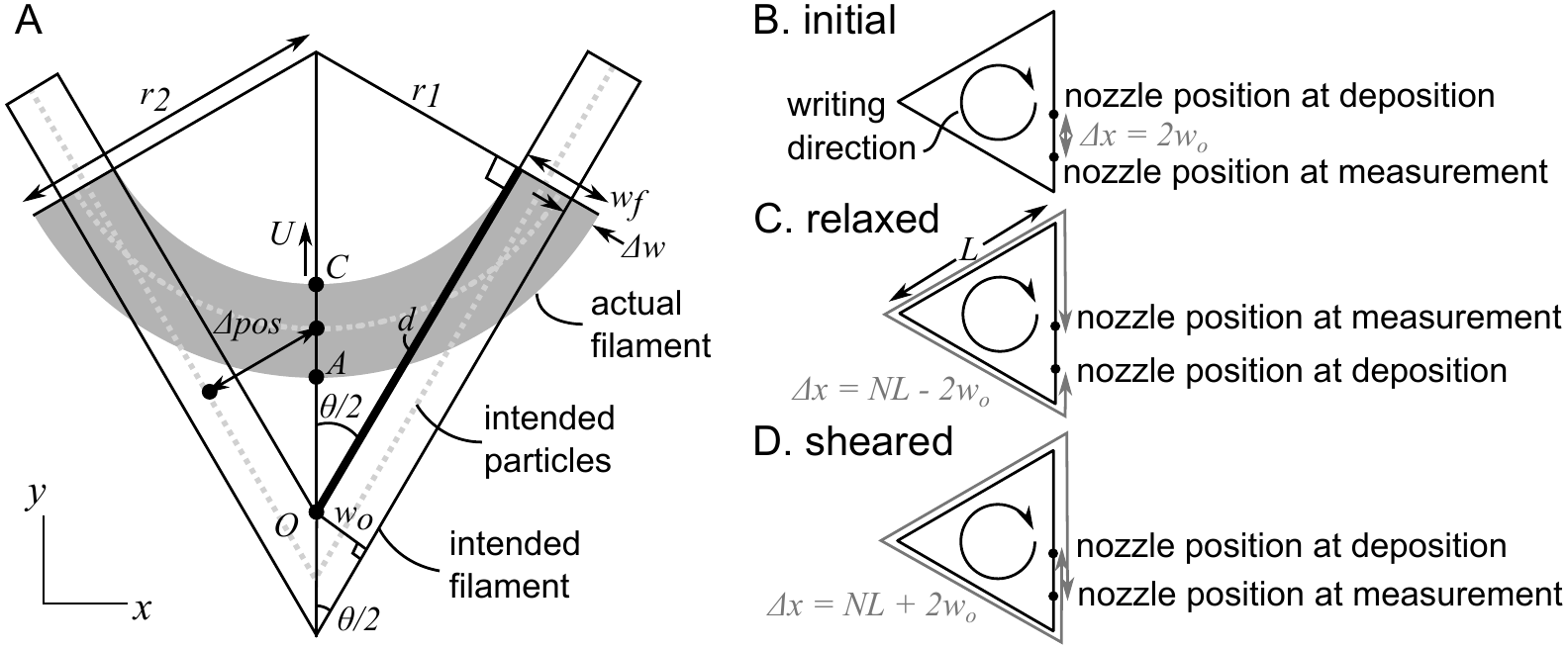}
    \caption[Schematic of Laplace pressure-driven smoothing]{Laplace pressure-driven smoothing. A) Schematic of variables used in derivation. The intended sharp corner (black outline) is replaced by a smooth arc (gray shaded area), shifting and widening the particle distribution. $d$ is the length of the bold line. B-D) Lengths traveled by the nozzle $\Delta x$ at the measurement of the initial, relaxed, and sheared distributions.}
    \label{fig:smoothschem}
\end{figure*}

In low-viscosity liquid elbows, capillarity drives fluid to flow toward the inner edge of the elbow due to a Laplace pressure differential between the inner and outer edge of the elbow that depends on the radii of curvature of the inner ($r_1$) and outer ($r_2$) surfaces of the elbow and the surface tension of the ink $\gamma$ (Fig. \ref{fig:smoothschem}A). The change in Laplace pressure from points A to C is given by

\begin{equation}
    \Delta P_{AC} = \gamma\Big(\frac{1}{r_1}+\frac{1}{r_2}\Big)
    \label{eq:laplace}
\end{equation}

One previously proposed model finds that an inviscid liquid elbow can achieve capillary equilibrium by forming a spherical sector-shaped bulge at the corner.\cite{Huang2017} For the inks used in this study, this bulge-based model predicts that a surface energy-driven energy barrier will suppress bulge formation (supplemental). Although the bulge-based model may contribute to some print speed-dependent effects to be discussed later, the printed geometries of the moderate viscosity inks used in this paper exhibit arcs at corners rather than bulges (Fig. \ref{fig:prints}). Thus, we propose an additional model for the position and width of the line based on the Laplace pressure differential, which drives movement of fluid from the outer edge of the corner toward the inner edge. The model uses viscous dissipation to determine the shape of the corner after a finite time. 

Assume that a designed infinitely sharp corner with corner angle $\theta$ is replaced by an arc of equal area to the intended filament (Fig. \ref{fig:smoothschem}A). The intended filament width is $w_o$. The arc has thickness $w_f$. The distance between the inner corner of the intended filament and the intersection between the inner edge of the arc and the inner edge of the intended filament is the displaced length $d$. To determine $\Delta{Width}$ and $\Delta{Position}$ given the corner angle $\theta$ and the initial line width $w_o$, we need to determine the displaced length $d$ and the arc thickness $w_f$. The arc thickness $w_f$ and the displaced length $d$ relate in such a way that $d$ and $w_f$ must be determined numerically using the equations listed in this section. 

From geometry,

\begin{equation}
    r_1 = d\tan\frac{\theta}{2}
    \label{eq:r1}
\end{equation}

and 

\begin{equation}
    r_2 = r_1+w_f
    \label{eq:r2}
\end{equation}

We can incorporate Equation \ref{eq:r1} and \ref{eq:r2} into Equation \ref{eq:laplace} to determine the Laplace pressure differential as a function of the displaced length $d$ and the arc thickness $w_f$.

\begin{equation}
    \Delta P(d, w_f) = \gamma\Big(\frac{1}{d\tan(\theta/2)}+\frac{1}{d\tan(\theta/2)+w_f}\Big)
    \label{eq:pdwf}
\end{equation}

Assume that point $C$ on the inner ink-substrate contact line advances toward the center of the arc with velocity $U$ (Fig. \ref{fig:smoothschem}A). Note that as point $C$ moves, the center of the arc also moves. The velocity is proportional to the Laplace pressure differential between the inner and outer surfaces of the corner. We can express $U$ as a function of the displaced length $d$:

\begin{equation}
    U(d, w_f) = \frac{\Delta P(d, w_f) \lambda}{\eta},
\end{equation}
where $\eta$ is the viscosity of the ink, and $\lambda$ is a length scale  along which the contact line is driven. $\lambda$ encompasses several thermodynamic parameters and molecular length scales.\cite{Blake2006} In this work, we empirically set $\lambda$ equal to 1 $\mu$m, which brings the predicted values to the same order of magnitude of the experimental values. Increasing $\lambda$ generally increases the predicted values of $\Delta Position$ and $\Delta Width$ uniformly across printing parameters (supplemental).

The distance between point $C$ and the inner point of the intended print path corner $O$ can be expressed in terms of $d$, as follows (Fig. \ref{fig:smoothschem}A):

\begin{equation}
    \overline{CO} = d\sec(\theta/2) - d\tan(\theta/2) = d\Big(\frac{1-\sin(\theta/2)}{\cos(\theta/2)}\Big)
    \label{eq:co1}
\end{equation}

Even though the contact line velocity depends on the displaced length $d$, assume that the contact line velocity over the entire smoothing process is constant and equal to the velocity at the final displaced length $d$. We can thus determine the distance that the contact line travels $\overline{CO}$ as a function of the contact line velocity.

\begin{equation}
    \overline{CO} = U(d, w_f)\frac{\Delta x}{v_s} = \frac{  \Delta x \Delta P(d, w_f) \lambda}{\eta v_s},
    \label{eq:co2}
\end{equation}
where $\Delta x$ is the length that the stage travels between deposition and measurement, and $v_s$ is the stage translation speed. By setting Equation \ref{eq:co1} equal to Equation \ref{eq:co2}, the displaced length $d$ can be determined for the traveled lengths $\Delta x$ of interest. 

The initial distribution is measured $\Delta x \approx2w_o$ behind the nozzle just after deposition (Fig. \ref{fig:smoothschem}B), so the displaced length $d$ for the initial distribution is

\begin{equation}
    d_{init} = \frac{2w_o\Delta P(d, w_f)\lambda}{\eta v_s}\frac{\cos(\theta/2)}{1-\sin(\theta/2)}
    \label{eq:dinit}
\end{equation}

The relaxed particle distribution is measured after traveling a length $\Delta x \approx NL - 2w_o$ around the polygon where $N$ is the number of sides on the polygon (3, 4, 5, 6, or 8 in these experiments), and $L$ is the polygon edge length (10 mm in these experiments)  (Fig. \ref{fig:smoothschem}C). The displaced length $d$ for the relaxed distribution is

\begin{equation}
    d_{relax} = \frac{(NL-2w_o)\Delta P(d, w_f)\lambda}{\eta v_s}\frac{\cos(\theta/2)}{1-\sin(\theta/2)}
    \label{eq:drelax}
\end{equation}

The sheared particle distribution is measured after traveling $\Delta x \approx NL + 2w_o$  (Fig. \ref{fig:smoothschem}C), so the displaced length $d$ for the sheared distribution is

\begin{equation}
    d_{shear} = \frac{(NL+2w_o)\Delta P(d, w_f)\lambda}{\eta v_s}\frac{\cos(\theta/2)}{1-\sin(\theta/2)}
    \label{eq:dshear}
\end{equation}

Note that the displaced lengths in Equations \ref{eq:dinit}--\ref{eq:dshear} depend on the arc thickness $w_f$. We can determine $w_f$ by setting the area of the arc equal to the area of the print path it is reforming.

%\begin{equation}
%    \frac{\pi-\theta}{2\pi}\pi(r_2^2 - r_1^2) = 2w_od+2\Big(1/2w_ow_o\cot\frac{\theta}{2}\Big)
%    \label{eq:smooth1}
%\end{equation}

\begin{equation}
    \frac{\pi-\theta}{2}(r_2^2 - r_1^2) = 2w_od+w_o^2\cot\frac{\theta}{2}
    \label{eq:smooth1}
\end{equation}

Substituting Equation \ref{eq:r1} and \ref{eq:r2} into Equation \ref{eq:smooth1} and solving the resulting quadratic for the arc thickness $w_f(d)$, one obtains:

\begin{equation}
    w_f(d) = \frac{-(\pi-\theta)d\tan(\theta/2) + \sqrt{(\pi-\theta)^2d^2\tan^2(\theta/2) - 2(\pi-\theta)(-2w_od-w_o^2\cot(\theta/2))}}{\pi-\theta}
    \label{eq:wfd}
\end{equation}

By combining Equation \ref{eq:wfd} with Equations \ref{eq:dinit}--\ref{eq:dshear} and Equation \ref{eq:pdwf}, we can numerically determine the value of the displaced length $d$ and the arc thickness $w_f$ in the initial measurement, after relaxation, and after shear. To evaluate these equations, we use the ink viscosity at a shear strain rate of 0.01 Hz, which is the lowest strain rate at which viscosities were measured (supplemental). We estimate surface tensions by using the surface tensions measured in Ref. \cite{Friedrich2018}, which use similar ink compositions, but with 16:84 wt silica:wt UDMA instead of 8:92 wt silica:wt UDMA. Because the surface tensions of the 16 wt\% silica inks are within 1 mJ/m\textsuperscript{2} of the surface tensions of inks without silica,\cite{Friedrich2018, Asmusen2009} we do not expect that the 8 wt\% difference in silica content will produce an appreciable difference in the estimated $\Delta Width$ and $\Delta Position$. 

Knowing $d$ and $w_f$, we can estimate $\Delta Width$ and $\Delta Position$. Because $w_f$ indicates the entire width of the printed filament, we can estimate the width of the particle distribution inside the printed filament by assuming that the particle distribution width is some fraction of the filament width. In this paper, the particle distribution width $\Delta Width$ is arbitrarily assumed to be one tenth of the printed filament width. A different assumption would scale all $\Delta{Width}$ values uniformly, but trends as a function of printing parameters would remain the same.

\begin{equation}
    \Delta{Width}_{smooth}(\theta, v_s, ink) = \frac{w_f-w_o}{10}
     \label{eq:wsmooth}
\end{equation}
 
From geometry, the change in the position of the center of the line (the particle distribution position) between the corner and center of the edge is

\begin{equation}
    \Delta Position_{smooth}(\theta, v_s, ink) = d\tan(\theta/2) + w_o/2 - (d\tan(\theta/2) + w_f/2)\sin(\theta/2)
    \label{eq:psmooth}
\end{equation}

Thus, from a corner angle $\theta$, ink composition, and translation speed $v_s$, we can predict the change in particle distribution position and width at the corner. We can compare these changes to measured experimental changes in particle distribution position and width at the corner. 

\subsection{Double deposition-driven swelling}

Geometric constraints imply that excess ink is deposited at corners during extrusion. Excess ink can come from two sources: double deposition and acceleration. Double deposition, where the nozzle must retrace some area on which it has already written, occurs in any scheme involving extrusion of filaments, including direct ink writing and fused deposition modeling.\cite{Han2002} Acceleration influences the corner shape at fast printing speeds. In these experiments, translation speeds are between 3 and 12 mm/s, which is slow enough that the 3-axis gantry used in this experiment accelerates over a trivial distance at corners. Specifically, the Shopbot ramp speed used here (the jerk speed, in extrusion-based 3D printing parlance) is greater than 12 mm/s, so linear speeds along the path center are not reduced before reaching the corner. A model that includes acceleration is described in the supplemental information. Here, we use a simplified analytical model to estimate how double deposition causes a change in distribution width and position at the corner.

\begin{figure}
    \centering
    \includegraphics{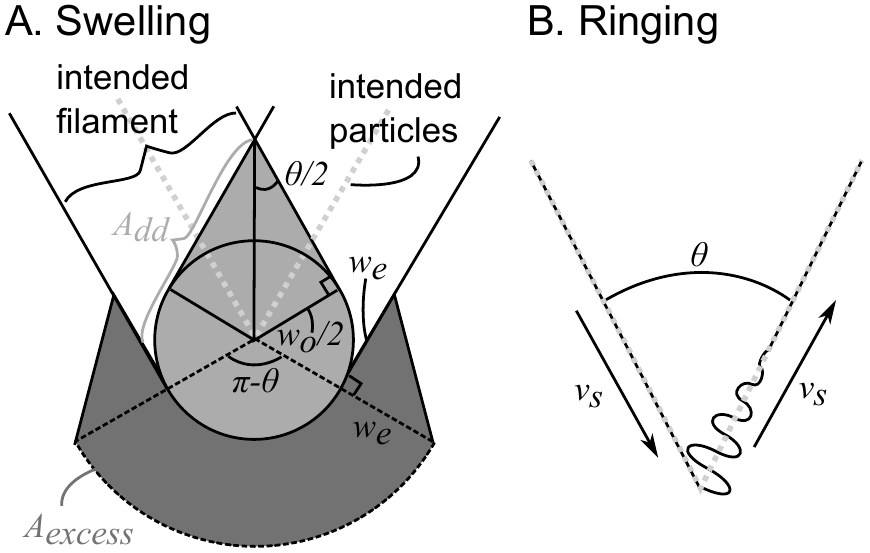}
    \caption[Schematic of double deposition-driven swelling and ringing]{A) Double deposition-driven swelling. B) Ringing due to fast turns. In this schematic, $v_s$ can be conceptualized as the nozzle translation speed.}
    \label{fig:theoryschems}
\end{figure}

Double deposition occurs because every time the print path changes direction, the nozzle retraces some area that it already covered. Conventionally, the double deposition area is calculated assuming that the nozzle is circular with radius $w_o/2$ (Fig. \ref{fig:theoryschems}A). If the nozzle stops moving at the corner, the resultant traced path will have an outer radius of $w_o/2$. In the experiments in this paper, the nozzle has a square cross-section, so the corner should match the square cross-section of the nozzle. However, the orientation of that square would vary based on the orientation of the corner. On average, we assume that the nozzle is circular with radius $w_o/2$. Also, we assume that the entire printed corner is of height $h$, so displaced volumes of ink are analogous to displaced areas of ink.

The the area of excess deposited ink is equal to the double deposition area $A_{dd}$, which can be expressed in terms of the corner angle $\theta$ and the corner width $w_o$.\cite{Han2002}

\begin{equation}
    A_{excess} = A_{dd} = \frac{w_o^2}{4}\Big(\cot\Big(\frac{\theta}{2}\Big) + \frac{\pi+\theta}{2}\Big)
    \label{eq:add}
\end{equation}

Ideally, the excess fluid would be deposited on the outer edge, within a sharp tip that falls within the intended print path.\cite{Han2002} However, the possibility of achieving such a sharp tip disagrees with numerical models and experimental results that indicate that the printed corner exhibits a rounded tip.\cite{Comminal2019, Kulkarni1999, Kao1998} Numerical models indicate that some of this excess volume will fall inside of the intended print path, and some will fall outside.\cite{Comminal2019} For a 90$^\circ$ corner, the volume of excess fluid deposited outside the print path corner is 2--3 times the volume deposited inside the corner.\cite{Comminal2019} This ratio expands to 10 times for 30$^\circ$ corners.\cite{Comminal2019} There is no accurate analytical model which predicts the ratio of excess volume deposited inside the corner to outside the corner. For simplicity, we assume that all of the excess fluid is deposited outside of the corner in an arc of thickness $w_e$, flanked by two triangles of height and width $w_e$ (Fig. \ref{fig:theoryschems}A). The material in these triangles helps to produce a smooth transition between the excess volume arc and the rest of the print path. Although the shape of this excess area contains unrealistically sharp edges, it approximates the shape of experimentally printed and numerically simulated corners.\cite{Comminal2019} Using geometry, the excess area depends on the arc width $w_e$:

\begin{equation}
    A_{excess} = \frac{\pi-\theta}{2}\Big(\Big(w_e+\frac{w_o}{2}\Big)^2-\Big(\frac{w_o}{2}\Big)^2\Big)+w_e^2
\end{equation}

With this, the arc width $w_e$ can be expressed in terms of the intended line width $w_o$, corner angle $\theta$, and excess corner area $A_{excess}$:

\begin{equation}
    w_e = \frac{-(\pi-\theta)w_o/2 +\sqrt{((\pi-\theta)w_o/2)^2 + 4((\pi-\theta)/2+1)A_{excess}}}{\pi-\theta+2}
    \label{eq:we}
\end{equation}

Equation \ref{eq:we} can be evaluated as a function of the intended line width $w_o$ and the corner angle $\theta$ by substituting Equation \ref{eq:add} into Equation \ref{eq:we}. 

The difference in line width between the corner and the intended line width is $w_e$. In this paper, the particle distribution width is assumed to be one tenth of the line width. Again, assuming a different particle distribution width ratio would not change the scaling of the change in width with corner angle, print speed, or ink viscosity. As such, the difference in particle distribution between the corner and center of the edge is: 

\begin{equation}
    \Delta Width_{swell}(\theta) = w_e/10
    \label{eq:wswell}
\end{equation}

Because the middle of the line and thus the particle distribution peak position shifts outward toward negative positions at the corner, the difference in the particle distribution peak position between the corner and center of the edge is

\begin{equation}
    \Delta Position_{swell}(\theta) = -w_e/2
    \label{eq:pswell}
\end{equation}

As the corner angle increases, the excess volume from double deposition decreases, so the magnitude of the change in position and width at the corner decreases.

Thus, we can predict the change in particle distribution position and width at the corner due to double deposition-driven corner swelling and compare these predicted values to measured experimental changes in the particle distribution position and width at the corner.

\subsection{Ringing}

If the translation speed is not reduced to zero at a corner, the rapid change in direction and speed that occurs at the corner induces vibrations in the 3-axis gantry.\cite{Comminal2019, Okwudire2018} These vibrations produce oscillatory deviations in the print path which effectively widen the printed line at the corner (Fig. \ref{fig:theoryschems}B). Because the oscillations are centered within the print path, the oscillations should not change the position of the line at the corner. 

\begin{equation}
    \Delta Position_{ring} = 0
    \label{eq:pring}
\end{equation}

Ringing should only impact the printed line coming out of the corner, not going into it (Fig. \ref{fig:theoryschems}B). Faster, sharper turns result in more ringing because they impose larger changes in speed over a small distance.\cite{Okwudire2018} Assume that ringing is extinguished at the center of the polygon edge. Assume that the amplitude of the oscillations and thus the width of the particle distribution at the corner scales with the change in velocity between the incoming and outgoing print paths:

\begin{equation}
    \Delta Width_{ring}(\theta, v_s) = C (2v_s\cos(\theta/2))
    \label{eq:wring}
\end{equation}

where $C$ is a damping factor that we empirically set to $C = 1/300$ seconds. Choosing a different value for $C$ does not impact the scaling of the change in width as a function of printing parameters and part design.

\section{Experimental approach}

This paper draws from the same data set that is used in Refs. \cite{Friedrich2019d, Friedrich2019s}, but this work only uses the first deposited line of the three-pass polygon, only uses 10 mm edge lengths, and focuses specifically on corners. Data and code can be found at Ref. \cite{Friedrich2019m}.

\begin{figure*}
	\centering\includegraphics[]{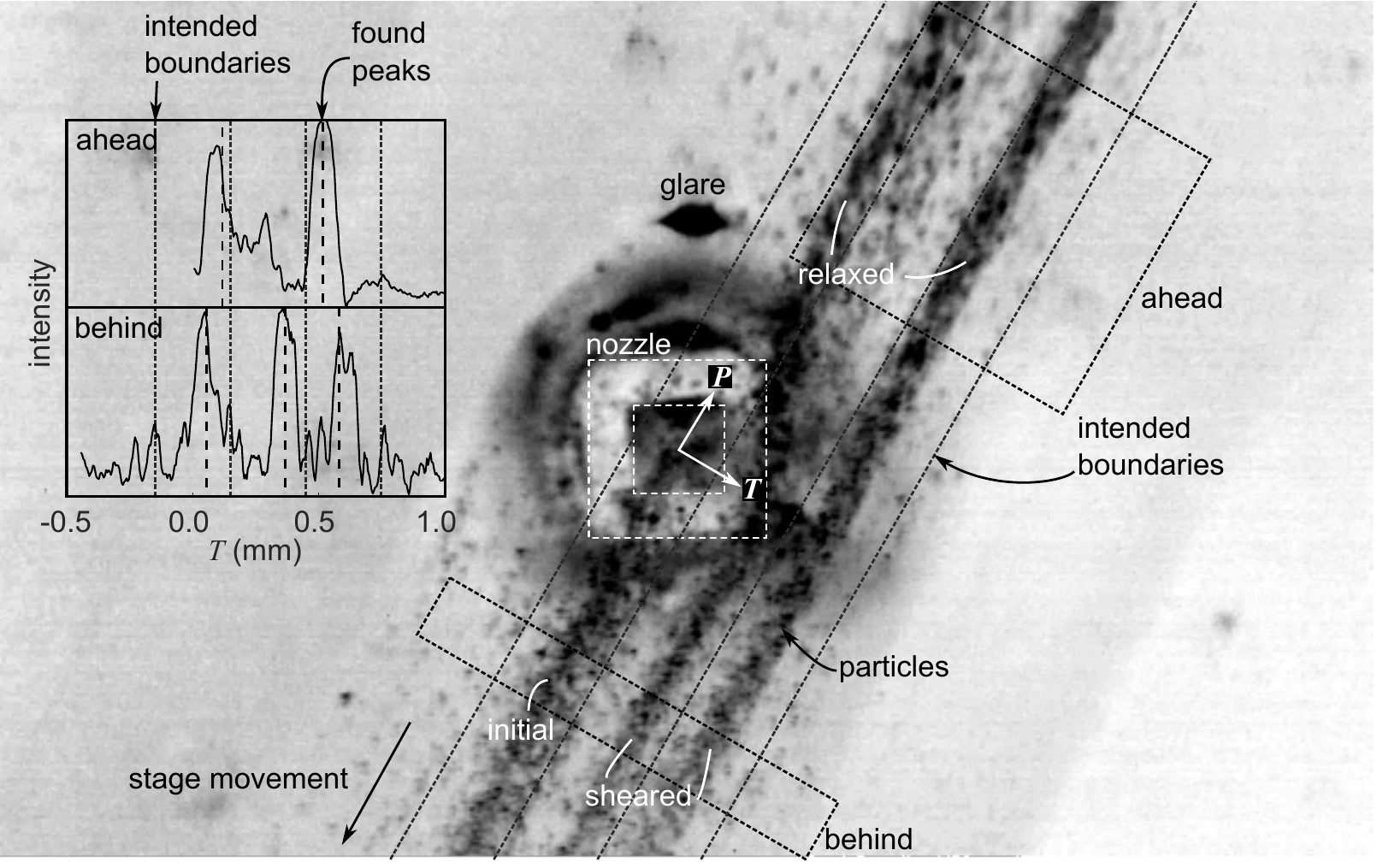}
	\caption[Example of analyzed frame.]{Example of analyzed frame, for layer-by-layer support. Image is inverted. Inset: Intensities are summed along the parallel direction ahead of and behind the nozzle. Distribution positions and widths are found as a function of transverse position.}
	\label{fig:methods}
\end{figure*}

\subsection{Materials}
Inks consisted of diurethane dimethacrylate (UDMA) (Sigma Aldrich, mixture of isomers with topanol inhibitor), triethylene glycol dimethacrylate (TEGDMA) (Sigma Aldrich, with MEHQ inhibitor), fumed silica (Evonik Aerosil R106), camphorquinone (CQ) (Sigma Aldrich), and 2-(Dimethylaminoethylmethacrylate) (DMAEMA) (Sigma Aldrich, with monomethyl ether hydroquinone inhibitor). Bases were mixed in a 92:8 UDMA:fumed silica weight ratio and mixed in a planetary mixer (Thinky ARE-310) at 2000 rpm for 3 minutes. Inks were then mixed in 80:20, 75:25, 70:30, and 65:35 base:TEGDMA weight ratios with 0.2 w\% CQ and 0.8 w\% DMAEMA, plus 10 w\% ($\approx$1.4 v\%) silver-coated copper microspheres (Potters Beads Conduct-O-Fil, SC15S15, diameter 15 $\mu$m), which were acoustically focused in the nozzle. Carbopol support gels were mixed by adding 1.2 w\% Carbomer 940 to deionized water (pH 3-4) and mixing with an overhead stirrer for 5 minutes at 1500 rpm or until dissolved. Gels were then neutralized using 50\% NaOH (Sigma Aldrich) and mixed in a planetary mixer (Thinky ARE-310) at 2000 rpm for 3 minutes to remove bubbles.

\subsection{Nozzle configuration and video collection}

Inks and support gels were extruded through square borosilicate capillary nozzles (Vitrocom, 0.3 mm ID, 0.6 mm OD, 50 mm length) onto glass slides. The ink capillary was seated in a stainless steel block containing a 0.7 mm square groove lined with ultrasonic coupling gel. A piezoelectric actuator (15 mm diameter x 1 mm thick Navy I material, American Piezo) was adhered to the steel block using epoxy (Devcon HP250). The piezo was driven using a signal generator (HP 33120A) and amplifier (Mini-Circuits LZY-22+), and signals were measured using an oscilloscope (Agilent DSO-X 2024A). Sinusoidal signals were generated at 2.249 MHz and a peak-to-peak voltage of 50 V\textsubscript{pp}. The piezo was thermally coupled to a copper cooling stage using thermal couplant (Wakefield type 120). Steel ferrules were bonded to capillaries using epoxy (Devcon HP250). Ink and support were extruded using a mass flow controller (Fluigent MFCS-EZ) at pressures calibrated by measuring masses extruded at fixed pressures and times. For ink, average flow speeds inside the nozzle were set equal to stage translation speeds.

The ink nozzle and substrate were illuminated using light transmitted through a red filter cube, to prevent curing. Videos were collected from underneath the nozzle through the glass substrate using a Point Grey Grasshopper GS3-U3-2356C-C camera with an Infinity Infinitube FM-200 objective and $\times$0.66 lens, at 86 fps.

For layer-by-layer support, the support gel was extruded through a square borosilicate capillary at a stage speed of 10 mm/s and an estimated flow speed of 15 mm/s. For bath support, a 1 mm-thick layer of support was spread onto the substrate. The point where the nozzle touches the substrate was defined as $z=0$. Parts were printed at a stand-off distance and line spacing of 0.3 mm, which is the inner width of the capillary. First, the three-line inner support polygon was printed from inside to outside, then the three-line outer support polygon from inside to outside, then the three-line ink polygon from inside to outside. Future experiments could print polygons from outside to inside to further examine the role of boundary conditions on corner defects. Equilateral triangles, squares, pentagons, hexagons, and octagons were printed with 10 mm edge lengths (Fig. \ref{fig:printdiag}). 

Distributions were collected using Matlab R2018b. Backgrounds were removed using a 15 px disk structuring element. Where the origin is at the nozzle center and the inner nozzle width is $w$, the largest fully imaged region upstream of the nozzle from $T = 0$ to $T = 3.5w$ and downstream of the nozzle from $T=-1.5w$ to $T = 3.5w$ were each summed along the print direction. Peaks were identified using the Matlab function \textit{findpeaks} with a minimum peak-to-peak distance of $w/2$. Frames in which the corner of the polygon is visible were removed. The width is the standard deviation of the distribution within $w/2$ on either side of the peak. Width and position measurements at the corners were taken from the starting and ending 2.5 mm of the 10 mm edge, and measurements at the center of the edge were taken from the middle 2 mm of the 10 mm edge (Fig. \ref{fig:timemaster}). The change in position and width at the corner are measured as the value at the corner, subtracted by the value at the center (Eq. \ref{eq:dpos1}, \ref{eq:dwidth1}).

\section{Results}

\begin{figure*}
    \centering
    \includegraphics{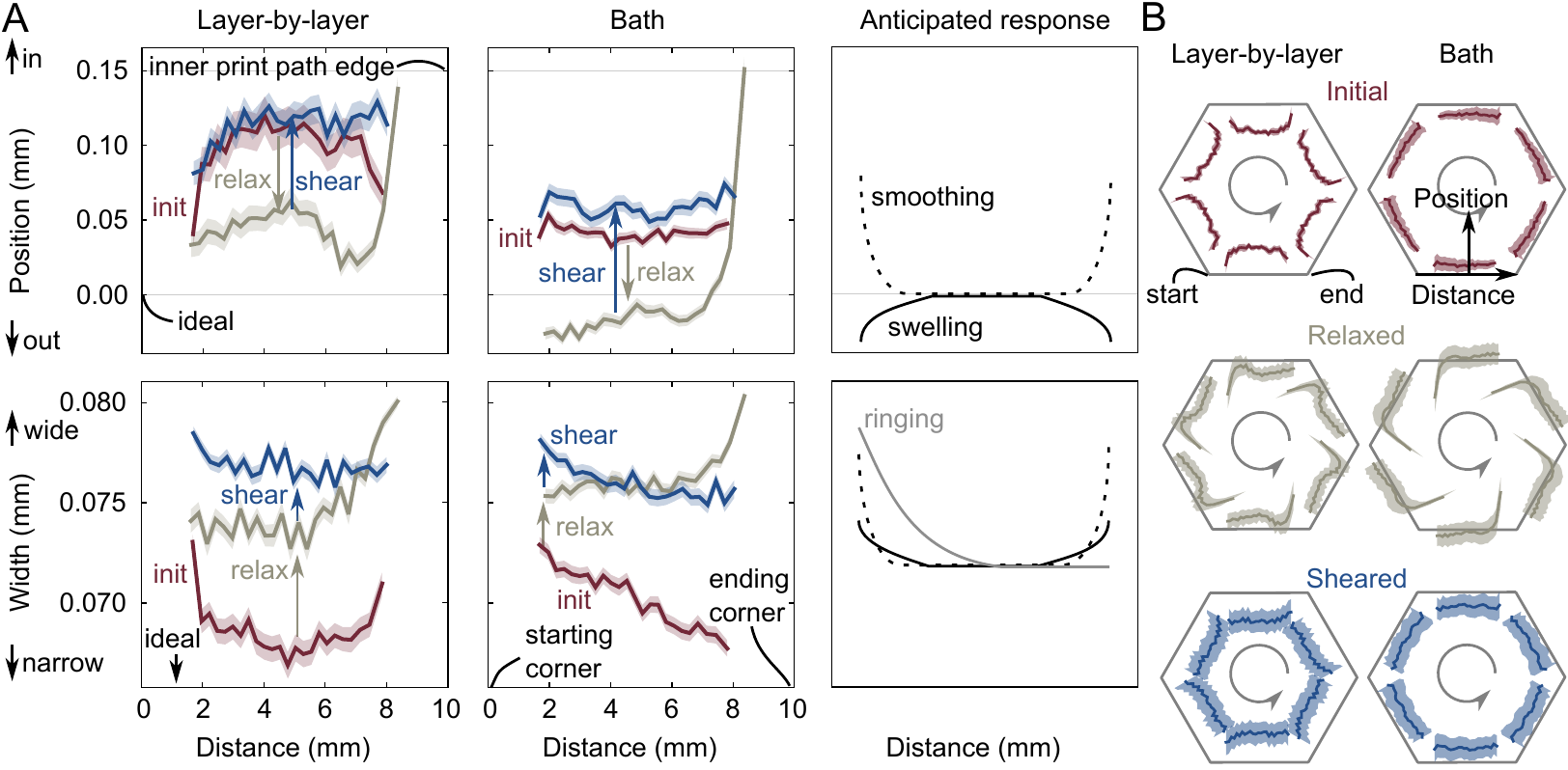}
    \caption[Particle distribution over length of the edge]{Shifts in particle distribution center and width over the polygon edge. A) Particle distribution position and width as a function of distance from the corner, for line 1 averaged over all edges, corner angles, print speeds, and ink compositions. Width is the standard deviation of the particle distribution. Shaded areas indicate standard error over many printing parameters. Anticipated response is arbitrarily scaled. B) Exaggerated illustrations of the initial, relaxed, and sheared particle distributions in printed hexagons. Dark lines represent the position, while light areas represent the distribution width. Because the scale has been amplified for visual comprehension, the jumps between the end of one edge and the start of the next in the relaxed state are not as jagged in reality.}
    \label{fig:timemaster}
\end{figure*}

Videos of the region near the nozzle during printing indicate that the behavior of the particle distribution at corners changes during the course of the print, and the nature of those changes depends on the geometry of the support material. In this section, an edge is a straight segment that connects a starting corner to an ending corner. Distances are measured from the starting corner (Fig. \ref{fig:timemaster}). Positive particle distribution positions are toward the inner edge of the polygon, while negative positions are toward the outer edge. Ideally, the change in the particle distribution position and width at the corner would be zero, ensuring consistency in microstructure throughout the print. In other words, the final location of the particle distribution is in the center of the designed print path, and the particle distribution is just as narrow at the corner as it is in the middle of a straight segment. Particle distributions are measured at three points in the printing process. (1) Corner defects can come from the deposition process and manifest in the initial distribution. (2) Defects can evolve over time during relaxation, so relaxed distributions are measured just before the next neighboring line is printed. (3) Finally, defects can appear when the nozzle returns to write a neighboring filament and shears the existing filament, so sheared distributions are measured just after the next neighboring line is printed. Initial and sheared distributions behave similarly, and relaxed distributions exhibit distinct trends. 

In this section, experimental changes in particle distribution at the corner $\Delta Position$ and $\Delta Width$ are measured. Trends in these values as a function of distance from the corner, corner angle, and print speed are compared to the theoretical contributions of swelling, smoothing, and ringing. Because all three theories were constructed with arbitrary scaling factors, only trends are compared between observations and predictions.

The variation in the particle distribution as a function of distance from the corner can be used to diagnose sources of corner defects. The change in particle distribution position $\Delta Position$ along the length of the edge varies based on support geometry. Figure \ref{fig:timemaster}A shows the particle distribution position and width as a function of distance from the starting corner. Figure \ref{fig:timemaster}B shows an exaggerated illustration of the particle distribution position and width on a printed hexagon to aid visualization. Smoothing causes inward shifts in position at corners, while swelling causes outward shifts at corners. In layer-by-layer support, the initial distribution shifts outward at corners, as predicted by swelling. In bath support, the initial distribution shifts slightly inward at corners, as predicted by smoothing. During relaxation, prior to the nozzle returning to write a neighboring line, the distribution changes. In layer-by-layer support, the relaxed distribution still bows inward at the center of the edge, but it sharply shifts inward at the ending corner, exhibiting traits of both swelling (at the center) and smoothing (at the corner). In bath support, although the distribution initially had an outward bow at the center of the edge, the distribution adopts a slight inward bow at the center during relaxation. Like layer-by-layer support, in bath support the relaxed distribution position sharply shifts inward at the ending corner, exhibiting traits of both swelling (at the center) and smoothing (at the corner). After the nozzle passes to write a new line, shearing the existing line, distributions once again resemble their initial positions. In layer-by-layer support, the sheared distribution shifts outward at corners, exhibiting traits of swelling. In bath support, the sheared distribution shifts inward at corners, exhibiting traits of smoothing.

Like the change in distribution position, the change in distribution width along the length of the edge also depends on support geometry (see results in Fig. \ref{fig:timemaster}). Smoothing and swelling cause the distribution to widen at both corners, while ringing only causes the particle distribution to widen at the starting corner. In layer-by-layer support, the initial distribution widens sharply at the starting and ending corners as predicted by the swelling and smoothing models, but in bath support, the initial distribution narrows gradually from the starting corner to the ending corner as predicted by the ringing model. After the line has had time to relax, in both support geometries the relaxed distribution widens sharply at the ending corner, amplifying an existing trend in layer-by-layer support but creating a new effect in bath support. After the nozzle returns to write a new line and shears the existing line, the sheared distribution width follows similar behaviors to the initial distribution. In layer-by-layer support, the sheared distribution only slightly widens at the starting corner but becomes roughly uniform along the edge length. Similarly, in bath support the sheared distribution narrows along the length of the edge as it did initially, but the sheared variation over the edge is less severe than the initial variation over the edge.

\begin{figure*}
    \centering
    \includegraphics{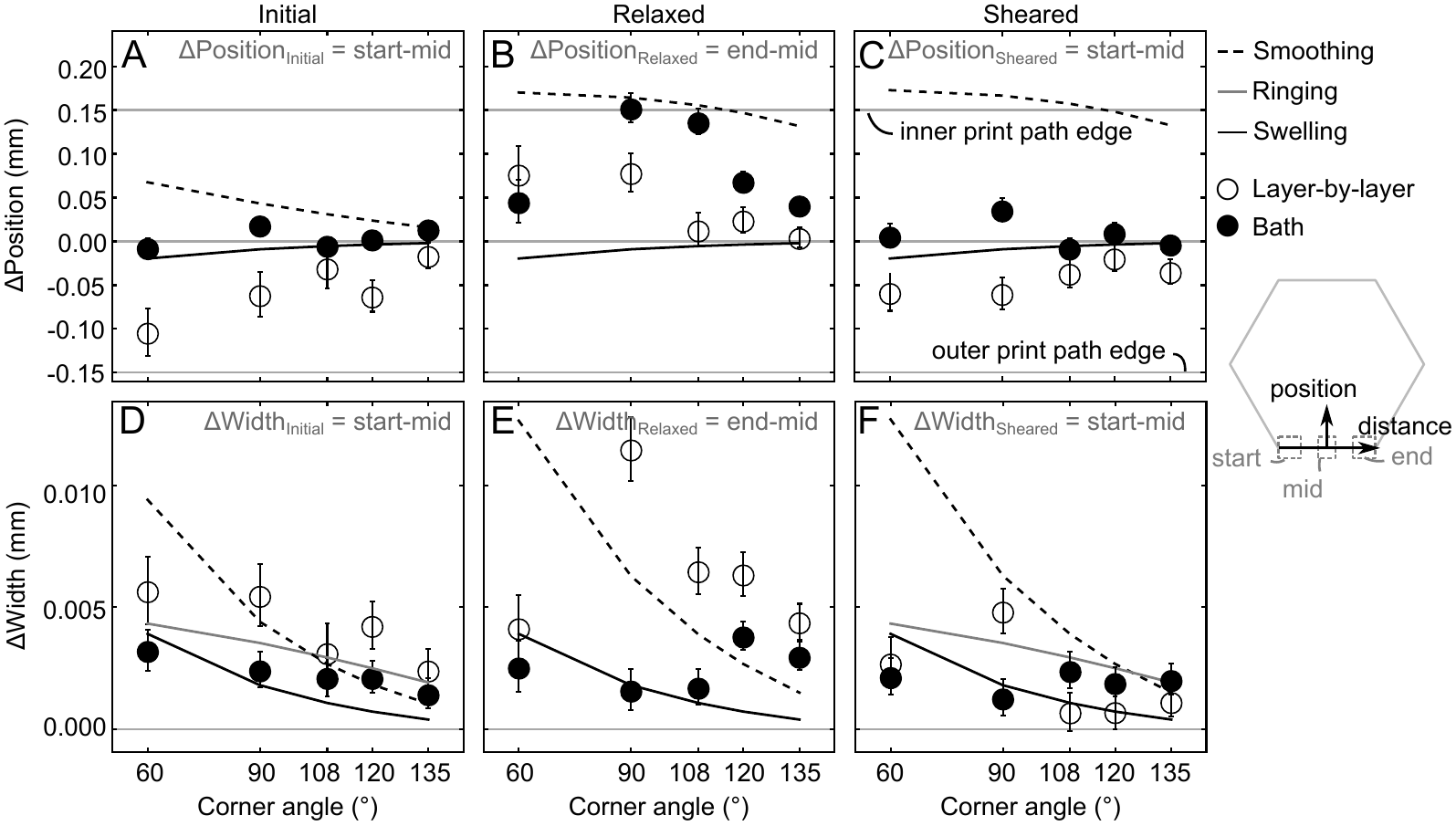}
    \caption[Corner defects as a function of corner angle]{Corner defect path shifts (A--C) and particle distribution spreading (D--F) at the corner as a function of polygon corner angle. Theoretical contributions of the smoothing, swelling, and ringing models are shown with continuous lines. Error bars indicate standard error.}
    \label{fig:timetheoryangle}
\end{figure*}

\begin{figure*}
    \centering
    \includegraphics{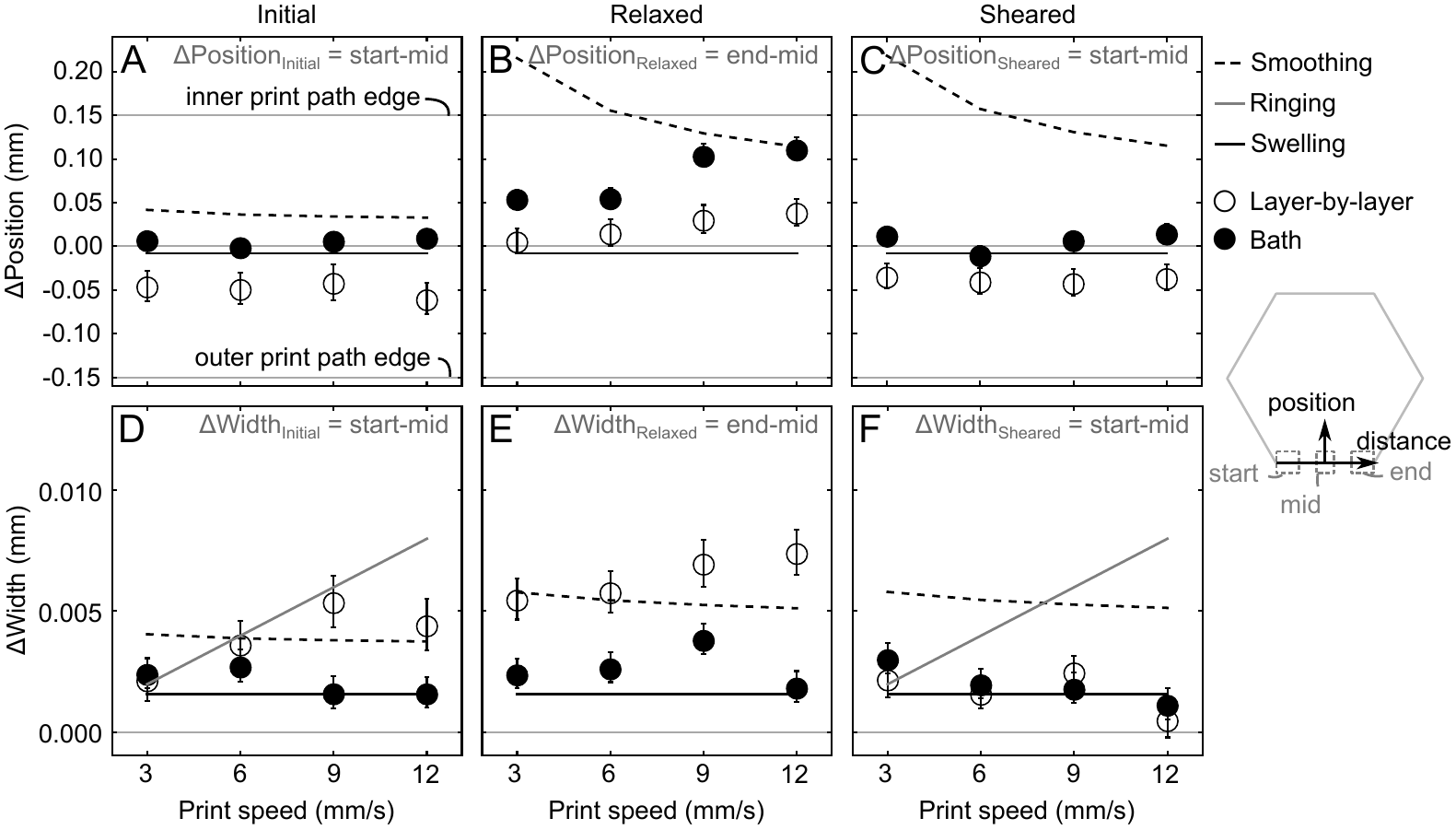}
    \caption[Corner defects as a function of print speed]{Corner defect path shifts (A--C) and particle distribution spreading (D--F) at the corner as a function of print speed, which represents both the translation and flow speeds. Theoretical contributions of the smoothing, swelling, and ringing models are shown with continuous lines. Error bars indicate standard error.}
    \label{fig:timetheoryv}
\end{figure*}

The difference in particle distribution position and width between the corner and middle of the polygon edge can further elucidate the sources of corner defects. Figure \ref{fig:timetheoryangle} shows these differences as a function of the corner angle, and Figure \ref{fig:timetheoryv} shows these differences as a function of print speed, which is equal to both the flow speed in the nozzle and the translation speed of the stage. Comparing experimental changes to the changes predicted in Equations \ref{eq:wsmooth}, \ref{eq:psmooth}, \ref{eq:wswell}, \ref{eq:pswell}, \ref{eq:pring}, and \ref{eq:wring} as shown by the continuous lines in Figure \ref{fig:timetheoryangle} and \ref{fig:timetheoryv} indicates which of the three mechanisms is the most likely source of corner defects. Note that there is no predicted relaxed change in width due to ringing because the relaxed change in width is measured between the ending corner and center of the edge, and ringing only impacts the starting corner. Because the initial and sheared distributions are measured behind the nozzle, differences are measured between the middle of the edge and the starting corner. Because the relaxed distribution is measured ahead of the nozzle, differences are measured between the middle of the edge and the ending corner.

The smoothing, swelling, and ringing models predict that the changes in the particle distribution at the corner vary with corner angle, ink composition, and print speed. Within the probed range, experimental effects of ink composition on changes at the corner are weak and may be found in the supplemental information. Trends as a function of corner angle and print speed are more conclusive.

The dependence of changes in the particle distribution at the corner on corner angle can be used to diagnose sources of microstructural defects at the corner. The initial and sheared changes in position $\Delta Position$ for layer-by-layer support are negative and decrease in magnitude with increasing corner angle, matching the swelling model (Fig. \ref{fig:timetheoryangle}A,C). In bath support, initial and sheared changes in position $\Delta Position$ are close to 0 for all corner angles, which could come from a combination of the two opposing mechanisms or from a suppression of both smoothing and swelling. For both supports, the relaxed change in position is positive and decreases in magnitude with increasing corner angle, matching the smoothing model (Fig. \ref{fig:timetheoryangle}B). 

The smoothing and swelling models both predict that the change in width at the corner $\Delta Width$ decreases with increasing corner angle. For both support geometries, the initial change in width decreases with increasing corner angle, matching both models (Fig. \ref{fig:timetheoryangle}D). In contrast, the relaxed change in width for bath support and sheared changes in width for both supports decrease, then increase with increasing corner angle, indicating that another effect may cause the particle distribution at the corner to widen at large corner angles during relaxation and possibly during shear (Fig. \ref{fig:timetheoryangle}E). 

The relaxed change in width $\Delta Width$ in layer-by-layer support (Fig. \ref{fig:timetheoryangle}E) and change in position $\Delta Position$ in both types of support (Fig. \ref{fig:timetheoryangle}B) decrease sharply from a corner angle of 90$^\circ$ to 135$^\circ$, but the changes are incongruously small at a corner angle of 60$^\circ$ for the position in bath support and width in layer-by-layer support. This may come from the design of the print path. The length of the polygon edge (10 mm) is established at the outer edge of the polygon, so the length of the inner edge of the polygon decreases at sharper corner angles, leading to smaller differences between the center and corner at 60$^\circ$. Alternatively, at a large spreading length scaling factor $\lambda$, the change in position due to corner smoothing decreases at 60$^\circ$, but not by as much as the experimental results (Fig. \ref{fig:lambdascale}).

The dependence of changes in the particle distribution at the corner on the print speed can be used to diagnose sources of microstructural defects at the corner. The smoothing model predicts that as the print speed increases, the change in position at the corner $\Delta Position$ should decrease (Fig. \ref{fig:timetheoryv}A--C). The swelling model predicts no print speed dependence because the model only considers double deposition, not acceleration at the corner. Experimental data indicate that for both supports, the initial and sheared changes in position do not change by more than one standard error across the tested print speeds, indicating that either the swelling model or no model dominates the initial and sheared change in position at the corner (Fig. \ref{fig:timetheoryv}A,C). The relaxed changes in position are positive and increase with print speed (Fig. \ref{fig:timetheoryv}B). Although the smoothing model predicts positive changes in position at the corner, it predicts that those changes should decrease with increasing print speed. As such, neither smoothing nor swelling describes the relaxed change in position as a function of print speed. 

The smoothing model predicts that the change in width $\Delta Width$ should slightly decrease with increasing print speed, the swelling model predicts no dependence on print speed, and the ringing model predicts that the change in width should increase with print speed (Fig. \ref{fig:timetheoryv}D--F). The initial and relaxed change in width are mostly invariant with print speed for bath support (Fig. \ref{fig:timetheoryv}D,E). The initial change in width increases with print speed for layer-by-layer support, as predicted by the ringing model (Fig. \ref{fig:timetheoryv}D). The relaxed change in width also increases with print speed in layer-by-layer support (Fig. \ref{fig:timetheoryv}E). This trend is not a residual effect from ringing in the initial change in width, because ringing occurs at the starting corner, and the relaxed change in width is measured at the ending corner. As such, none of the discussed theories explain the increase in relaxed change in width with print speed. Sheared changes in width at the corner decrease with increasing print speed for both supports, qualitatively matching the smoothing model but exhibiting a more severe dependence on print speed than predicted by the smoothing model (Fig. \ref{fig:timetheoryv}F). 

\section{Discussion}

A summary of the instances where the experiments qualitatively match the proposed theories is shown in Table \ref{tab:sumtable}. Because smoothing due to capillarity is hindered by viscous dissipation, smoothing should be influential on long time scales, after relaxation. Swelling should act on short time scales and be visible in the initial distribution because it is not hindered by viscous dissipation, and excess fluid must be deposited somewhere immediately. Similarly, ringing should act immediately and be visible in the initial distribution because it is directly changes the path along which the stage travels. In Table \ref{tab:sumtable}, no single theory explains every behavior of the initial, relaxed, or sheared distribution in layer-by-layer or bath support. Rather, various theories match the various experimental behaviors, suggesting that several mechanisms influence the particle distribution at each point in the printing process. Further, because some behaviors are not explained by any of the considered mechanisms, it is possible that the process is also influenced by other driving forces. 

\begin{table*}
    \centering
    \begin{tabular}{l|lll|lll}
        &\multicolumn{3}{c|}{Layer-by-layer} & \multicolumn{3}{c}{Bath}\\
        & Distance & Corner angle & Print speed & Distance & Corner angle & Print speed  \\
        \hline
        Initial position & \textbf{sw} & \textbf{sw} & \textbf{sw} & \textcolor{Maroon}{sm} & \textbf{sw}+\textcolor{Maroon}{sm} or none & \textbf{sw} or none \\
        Initial width & \textcolor{Maroon}{sm} or \textbf{sw} & any & \textbf{\textcolor{Gray}{ri}} & \textbf{\textcolor{Gray}{ri}} & any & \textbf{sw} or none \\
        \hline
        Relaxed position & \textcolor{Maroon}{sm}+\textbf{sw} & \textcolor{Maroon}{sm} & ? & \textcolor{Maroon}{sm}+\textbf{sw} & \textcolor{Maroon}{sm} & ?\\
        Relaxed width & \textcolor{Maroon}{sm} or \textbf{sw} & ? & ? & \textcolor{Maroon}{sm} or \textbf{sw} & ? & \textbf{sw} or none \\
        \hline
        Sheared position & \textbf{sw} & \textbf{sw} & \textbf{sw} & \textcolor{Maroon}{sm} & \textbf{sw}+\textcolor{Maroon}{sm} or none & \textbf{sw} or none \\
        Sheared width & none & ? & \textcolor{Maroon}{sm} & \textbf{\textcolor{Gray}{ri}} & ? & \textcolor{Maroon}{sm}\\ 
    \end{tabular}
    \caption[Comparison of theory and experiment]{Summary of cases where theory fits the experiments, in terms of the position or width as a function of distance or the trend in $\Delta$position or $\Delta$width as a function of corner angle or print speed. Experiments match the smoothing (\textcolor{Maroon}{sm}), ringing (\textbf{\textcolor{Gray}{ri}}), and/or swelling (\textbf{sw}) theory, exhibit trends that could come from a combination of theories or from the absence of all effects (none), or exhibit trends that occur in none of the theories (?).}
    \label{tab:sumtable}
\end{table*}

As shown in Table \ref{tab:sumtable}, the initial change in the distribution at the corner is dominated by swelling and ringing. Swelling is especially prominent in layer-by-layer support. The data leave some ambiguity as to whether layer-by-layer or bath support is more susceptible to ringing. The prominence of swelling and ringing in the initial distribution supports the hypothesis that swelling and ringing should act on short timescales. In bath support, changes at the corner in the initial distribution are mild. One explanation for these small changes is that smoothing and swelling are both present and cancel each other out. A more likely explanation is that both swelling and smoothing are suppressed in the initial distribution in bath support. The suppression of swelling and smoothing in bath support may come from hydrostatic stress. In layer-by-layer support, the inner edge of the corner is supported by support material, and the outer edge is exposed to air (Fig. \ref{fig:printdiag}). As such, the support material prevents deposition of excess volume on the inner edge, so all excess fluid should be deposited on the outer edge. Moreover, because there is only air on the outer edge of the corner, the excess fluid is capable of concentrating into more of a bulge as predicted by the model proposed by Huang, et. al.\cite{Huang2017} In contrast, in bath support, the corner is supported on the inner and outer edges by support material, so there is less of a preference for excess fluid to be deposited on the outer edge than in layer-by-layer support. Furthermore, because the corner is supported on its outer edge by support material, the hydrostatic stress from the support material may spread the excess ink out onto a long arc of the corner (i.e. longer triangles in the swelling model in Figure \ref{fig:theoryschems}A) rather than allowing the ink to concentrate into a capillarity-driven bulge.

Smoothing occurs during relaxation for both support types, which supports the hypothesis that smoothing acts on longer time-scales because of viscous dissipation (Table \ref{tab:sumtable}). After relaxation, particle distribution positions are still shifted inward at the center of the edge, exhibiting residual effects from swelling. Very close to the corner the particle distribution sharply shifts inward, matching smoothing (Fig. \ref{fig:timemaster}). This suggests that smoothing influences the position within a smaller distance from the corner than swelling but is capable of creating larger shifts in position than swelling. Layer-by-layer support produces smaller relaxed changes in position at the corner and larger changes in width at the corner than bath support. The difference in the change in position between the relaxed and initial states is roughly the same for layer-by-layer and bath support. As such, the difference in relaxed change in position between bath and layer-by-layer support is likely a residual difference from the initial distribution, where the change in position in bath support was already more positive than the change in layer-by-layer support. Because both support geometries support the inner edge of the corner, both types of support experience the same hydrostatic resistance to Laplace pressure-driven smoothing. Although the surface energy at the outer edge of the layer-by-layer supported corner is different from the surface energy at the outer edge of the bath supported corner, this difference in surface energy does not appear to produce a large difference in smoothing between the support types.

The sheared change in position and width at the corner largely follow the swelling model, like the initial change in position and width. This match in trends is misleading: swelling should occur instantly at the point of deposition and should not occur again in an already-deposited line. It is more useful to consider the forces imposed by the nozzle that has returned to print a new line. For the initial and relaxed distribution, changes at the corner occur largely because the corners shift more than the center. In contrast, in the sheared distribution the center shifts more than the corners. Consider Figure \ref{fig:methods}. The nozzle is moving toward the top right corner of the image, and the two relaxed lines are shifting toward the center of the polygon, in the bottom right corner of the image. During relaxation, the entire line shifts outward toward negative positions (Fig. \ref{fig:timemaster}). During shear, the nozzle shifts the existing line inward toward positive positions and away from the nozzle. The corners shift inward by less than the center because the corners are farther away from the nozzle due to capillarity-driven smoothing during relaxation. As such, although the sheared change in position at the corner exhibits some characteristics of both smoothing and swelling, we expect that neither smoothing or swelling is directly in control of the change in distribution between the sheared and relaxed states. We can use this framework of viewing the nozzle as a shearing force to understand why the sheared change in width at the corner does not strongly decrease as a function of corner angle, despite the predictions of all three models. The framework also helps to explain why the change in width at the corner decreases more steeply as a function of print speed than predicted by smoothing. As the nozzle deposits a new line on the outer perimeter of the existing line, the new line swells and rings at the corners, driving that new excess volume into the existing line and narrowing the existing line at the corner. Effectively, ringing and swelling in a neighboring line attenuate the effects of ringing and swelling in the original line. 

This result is useful because it implies that the final particle distribution at the end of printing should not vary greatly between corners of different angles, ensuring consistent properties throughout the print. Moreover, the sheared distribution is the best indicator of the final distribution of particles in the printed structure. Ideally, the change in position and change in width at the corner would be zero, ensuring a consistent distribution of particles throughout the entire print. Bath support achieves the smallest sheared change in position, and the sheared change in width is roughly the same between the two support geometries. As such, bath support should produce more consistent particle distributions than layer-by-layer support in structures printed with DIW with acoustophoresis.

The changes in relaxed position and width at the corner exhibit unexpected trends in print speed. Whereas the smoothing model predicts a small change and the swelling model predicts no change at higher print speeds, experimental data show large changes for both at faster print speeds. We do not expect ringing to directly explain this effect because ringing is caused by vibrations which occur just after the printer changes direction, so oscillations in the print path should only occur at the starting corner. The relaxed distribution is measured at the ending corner. Instead, faster print speeds may enhance the effects of capillarity. Specifically, increasing the flow speed and the translation speed increases the shear strain rate on the ink, lowering its viscosity and accelerating Laplace pressure-driven corner smoothing. Alternatively, the higher shear strain rate and larger vibrations in the moving stage at fast translation speeds may provide sufficient energy to overcome the energy barrier that suppresses capillarity-driven bulge formation as predicted by the model proposed by Huang, et. al. (see supplemental).\cite{Huang2017} The formation of a bulge in the inner edge of the corner would widen the distribution and shift it inward, matching the trend shown in the relaxed distribution (Fig. \ref{fig:timetheoryv}). However, the arc-shaped geometry exhibited by the corners printed in this study does not match the geometry predicted by the bulge-based model, so a model that combines the Huang, et. al. model with the model proposed in this paper may match these experiments more comprehensively.

The data do not convey a conclusive argument as to which support geometry is more vulnerable to ringing. Two trends convey opposing narratives. First, in bath support the initial and sheared particle distribution gradually narrows over the length of the edge as predicted by ringing, while in layer-by-layer support the particle distribution widens sharply at the starting and ending corners as predicted by smoothing and swelling, suggesting that bath support is more vulnerable to ringing. Second, ringing causes the initial change in width at the corner to increase with increasing print speeds, which is prominent for layer-by-layer support but not present in bath support, suggesting that layer-by-layer support is more vulnerable to ringing. One could argue that bath support should be more impacted by ringing because it more precisely preserves the line as written, whereas layer-by-layer support allows the line to blur more, allowing the other effects to overpower ringing. Alternatively, one could argue that bath support provides extra damping that prevents vibrations from impacting the printed line because the nozzle is fully submerged in bath support, so the stage is more closely coupled to the nozzle in bath support than in layer-by-layer support. Because of these two opposing trends and because there are some gaps in explanation in Table \ref{tab:sumtable}, it is possible that we have mischaracterized these experimental trends as ringing, and the trends may come from another source we have not considered.

\section{Conclusions}

During deposition of lines containing a narrow distribution of particles via direct ink writing with acoustophoresis, three effects cause changes in the particle distribution at printed corners. First, capillarity causes the corner to widen and travel inward, via corner smoothing. Corner smoothing occurs on long timescales and manifests after the printed corner has had time to relax. Second, the change in direction at the corner causes the nozzle to retrace a double deposition area. The excess volume from this double deposition is largely diverted to the outer edge of the corner, via corner swelling. Corner swelling occurs quickly during deposition and manifests in the initially deposited line. Third, when a 3-axis gantry changes direction without decelerating into the corner, the printer vibrates and traces an oscillatory path as it exits the corner, effectively widening the distribution of particles, via ringing. Ringing occurs quickly during deposition and manifests in the initially deposited line. In other moderate-viscosity direct ink writing applications without acoustophoresis or support material, we expect all three of these effects to still influence the shape and position of printed corners. 

There is much evidence of corner swelling in the initial and sheared particle distribution, particularly in layer-by-layer support. There is also evidence of smoothing in both types of support in the relaxed distribution, but some trends in the relaxed distribution remain unexplained. It is possible that another mechanism that strongly increases with print speed also induces corner defects during relaxation. Alternatively, the formulation of the smoothing model without support material could lead to an inaccurate scaling. Faster print speeds could increase the shear strain rate on the viscoelastic support material, yielding a larger volume of support and enabling smoothing to progress over a longer timescale. Finally, although there is strong evidence of swelling in the sheared particle distribution position, there is less evidence of swelling in the sheared distribution width. This is likely because corner defects in sheared distributions are controlled by the behavior of neighboring lines. The sheared corner doesn't swell again; the new corner swells and causes the existing corner to shift along with it, as demonstrated in the straight segments of polygons.\cite{Friedrich2019s} In straight segments, sheared distribution widths do not correlate with initial distribution widths, which may explain why only sheared positions match the swelling model, not sheared widths.

To ensure consistent properties throughout the print, it is necessary to limit changes in the particle distribution at corners and to limit differences in the particle distribution among corners of different angles. Some of these effects may be remediated using \textit{in-situ} curing. The inks used in this work are photopolymers, so a curing lamp in the direct-write printer could preserve printed microstructures at chosen points in the printing process.

\begin{itemize}
    \item To suppress particle positioning defects at corners, use a support bath, and design toolpaths with obtuse corners.
    \item To limit the effects of swelling and ringing, cure the deposited structure only after relaxation, because relaxation offsets their effects.
    \item To limit the effects of smoothing, cure the deposited structure just after deposition or just after the whole layer is complete.
\end{itemize}

\section*{Conflicts of interest}
There are no conflicts to declare.

\section*{Acknowledgments}
This work was supported by the Institute for Collaborative Biotechnologies through contract no. W911NF-09-D-0001 from the U.S. Army Research Office and a UCSB Chancellor's Fellowship. The work used the Microfluidics Laboratory at the California Nanosystems Institute and the Polymer Characterization Facility supported by the MRSEC Program of the National Science Foundation under award NSF DMR 1121053 at UCSB.

\newpage

\bibliography{ms}
\bibliographystyle{plainnat}

\section{Supplemental information: Corner accuracy in direct ink writing with support material}

 \renewcommand{\thesection}{S\arabic{section}}
 \renewcommand{\thefigure}{S\arabic{figure}}
 \renewcommand{\thetable}{S\arabic{table}}
 \renewcommand{\theequation}{S\arabic{equation}}
 
 \section{Theory}
 
 \begin{figure}[H]
 	\centering
 	\includegraphics{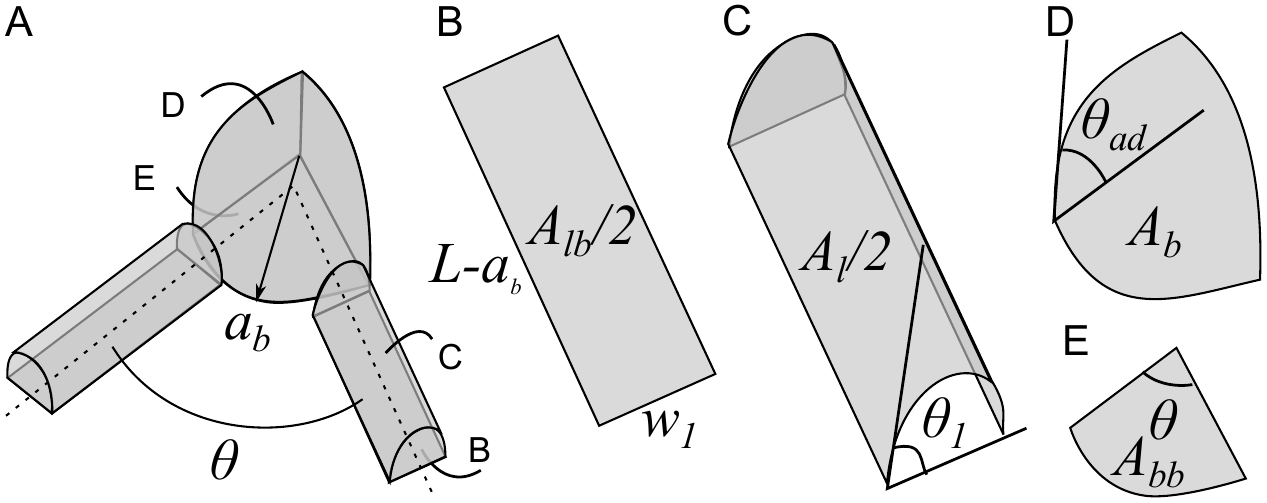}
 	\caption{Geometry of bulge-based corner smoothing model. A) 3D model of bulge and straight lines. B) Bottom surface of straight line. C) Top surface of straight line. D) Top surface of bulge. E) Bottom surface of bulge.}
 	\label{fig:huangmodel}
 \end{figure}
 
 The bulge-based corner smoothing model described here is adapted from Ref. \cite{Huang2017}. The model assumes that at a liquid elbow between two lines with circular segment cross-sections will split into three shapes: two shorter lines with circular segment cross-sections and one bulge shaped as a spherical sector. The total energy of the system $E_{tot}$ can be determined from the interfacial energies $\gamma$ and interfacial areas $A$:
 
 \begin{equation}
 E_{tot} = \gamma_{LG} (A_b + A_l) + (\gamma_{SL} - \gamma_{SG}) (A_{bb}+A_{lb})
 \label{eq:etot1}
 \end{equation}
 
 where $LG$ is ink-gas, $SL$ is substrate-ink, and $SG$ is a substrate-gas interface. The equilibrium contact angle $\theta_{eq}$ provides a relationship between the surface energies:
 
 \begin{equation}
 \gamma_{LG}\cos\theta_{eq} + \gamma_{SL} = \gamma_{SG}
 \label{eq:eqtheta}
 \end{equation}
 
 Substituting Equation \ref{eq:eqtheta} into Equation \ref{eq:etot1}, we find the total energy normalized by the ink surface energy $\gamma_{LG}$:
 
 \begin{equation}
 \frac{E_{tot}}{\gamma_{LG}} = A_b + A_l - \cos\theta_{eq}(A_{bb} + A_{lb})
 \label{eq:eqtot2}s
 \end{equation}
 
 From geometry, we can determine the areas $A_b$, $A_l$, $A_{bb}$, and $A_{lb}$.
 
 The width of the lines is $w_1$. Because our nozzle inner width is 0.3 mm, we set this value to 0.3 mm.
 \begin{equation}
 w_1 = 0.3 \quad\text{mm}
 \end{equation}
 
 The total volume of the system is the area of the nozzle ($w_1^2$) multiplied by the length of the elbow, which we set here to half the length of a polygon edge.
 \begin{equation}
 V_{tot} = 2L w_1^2
 \end{equation}
 \begin{equation}
 L = 5 \quad\text{mm}
 \end{equation}
 Given a bulge radius $a_b$, the area of the surface of the bulge $A_b$ depends on the advancing contact angle $\theta_{ad}$.
 \begin{equation}
 A_b = \frac{\theta}{2\pi}\pi a_b^2\Big(1+\tan^2\Big(\frac{\theta_{ad}}{2}\Big)\Big)
 \end{equation}
 
 The area of the base of the bulge $A_{bb}$ is
 \begin{equation}
 A_{bb} = \frac{\theta}{2\pi}\pi a_b^2
 \end{equation}
 
 The volume of the bulge $V_b$ is 
 \begin{equation}
 V_b = \Bigg(\frac{\theta}{2\pi}\Bigg)^{3/2}\Bigg(\frac{\pi\tan(\theta_{ad}/2)(3+\tan^2(\theta_{ad}/2))}{6}\Bigg)a_b^3
 \end{equation}
 
 The volume of the lines $V_l$ is
 \begin{equation}
 V_l = V_{tot}-V_b
 \end{equation}
 
 From the volume of the lines, the contact angle of the lines can be determined. We solve this equation using the Mathematica function FindRoot.
 \begin{equation}
 \frac{\theta_1-\sin\theta_1\cos\theta_1}{\sin^2\theta_1} == \frac{V_1}{2L (w_1/2)^2}
 \label{eq:theta1}
 \end{equation}
 
 The area of the base of the lines is
 \begin{equation}
 A_{lb} = 2w_1(L-a_b) 
 \end{equation}
 and the area of the surface of the lines is
 \begin{equation}
 A_l = \frac{V_l}{w_1/2}\Big(\frac{2\theta_1\sin\theta_1}{\theta_1 - \sin\theta_1\cos\theta_1}\Big)
 \end{equation}
 
 Given a corner angle $\theta$, a polygon edge length $2L$, an advancing contact angle $\theta_{ad}$, and an equilibrium contact angle $\theta_{eq}$, the equilibrium radius of the bulge is the radius $a_b$ that minimizes the energy determined in Equation \ref{eq:eqtot2}. We determine this value numerically for each combination of ink and corner angle using the contact angles in Table \ref{tab:contactangles}, which were measured using sessile droplets and the submerged needle technique. 
 
 \begin{table}[H]
 	\centering
 	\begin{tabular}{rccc}
 		w\% TEGDMA & $\theta_{ad}$ ($^\circ$) & $\theta_{eq}$ ($^\circ$) \\
 		20 & 58.59 (0.67) & 42.50 (0.59) \\
 		25 & 59.05 (1.23) & 37.23 (0.64) \\
 		30 & 49.83 (1.19) & 34.50 (0.82) \\
 		35 & 50.52 (1.20) & 32.68 (1.20) \\
 	\end{tabular}
 	\caption[Contact angles]{Advancing and static contact angles for inks with varying TEGDMA loadings. Parentheticals indicate standard error.}
 	\label{tab:contactangles}
 \end{table}
 
\begin{figure}
 	\centering
 	\includegraphics{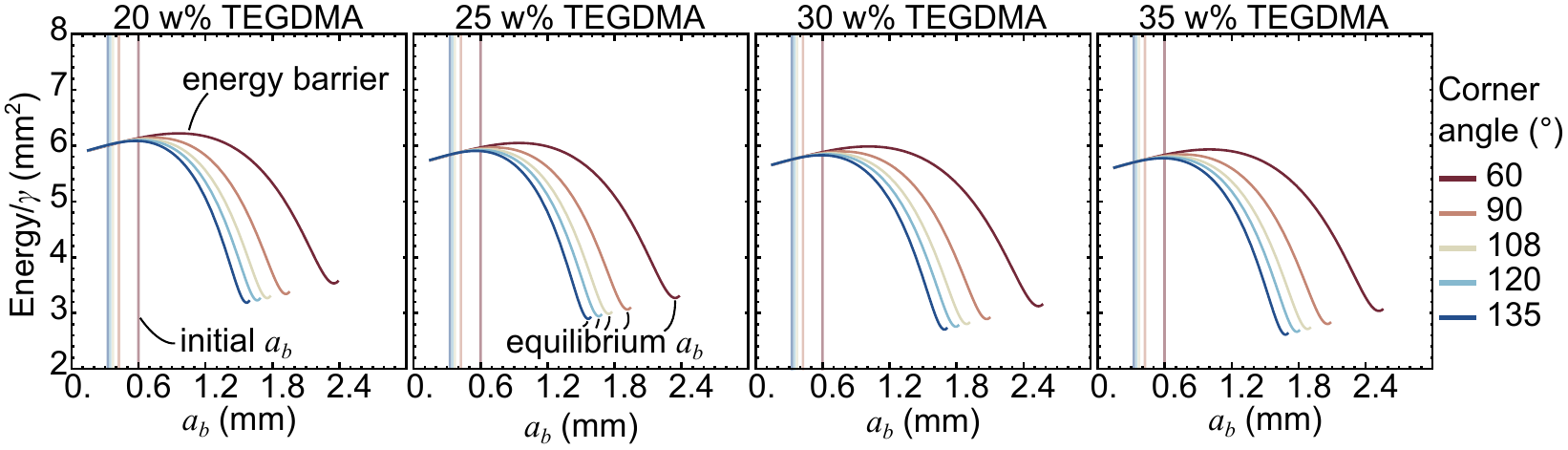}
 	\caption[Energy barrier for bulge formation]{Energy per surface tension $E_{tot}/\gamma_{LG}$ as a function of bulge radius $a_b$, for corner angles and inks used in this study.}
 	\label{fig:eqab}
 \end{figure}

 We expect the equilibrium measured radius of the bulge to be $a_b$ for initial, relaxed, and sheared distributions. However, for the inks used in this study, the energy curve from Equation \ref{eq:eqtot2} has an energy barrier (Fig. \ref{fig:eqab}). The initial radius $a_b = w_1/\sin(\theta/2)$ is less than the radius at the energy barrier. Because the inks are supported on their inner edge by a viscous Bingham plastic support material, we do not expect that enough energy will be supplied to the system to overcome the energy barrier, so bulges will not form. As such, we use a different model to predict the effect of capillarity, described in the main body of the paper.
 
 \begin{figure}
 	\centering
 	\includegraphics{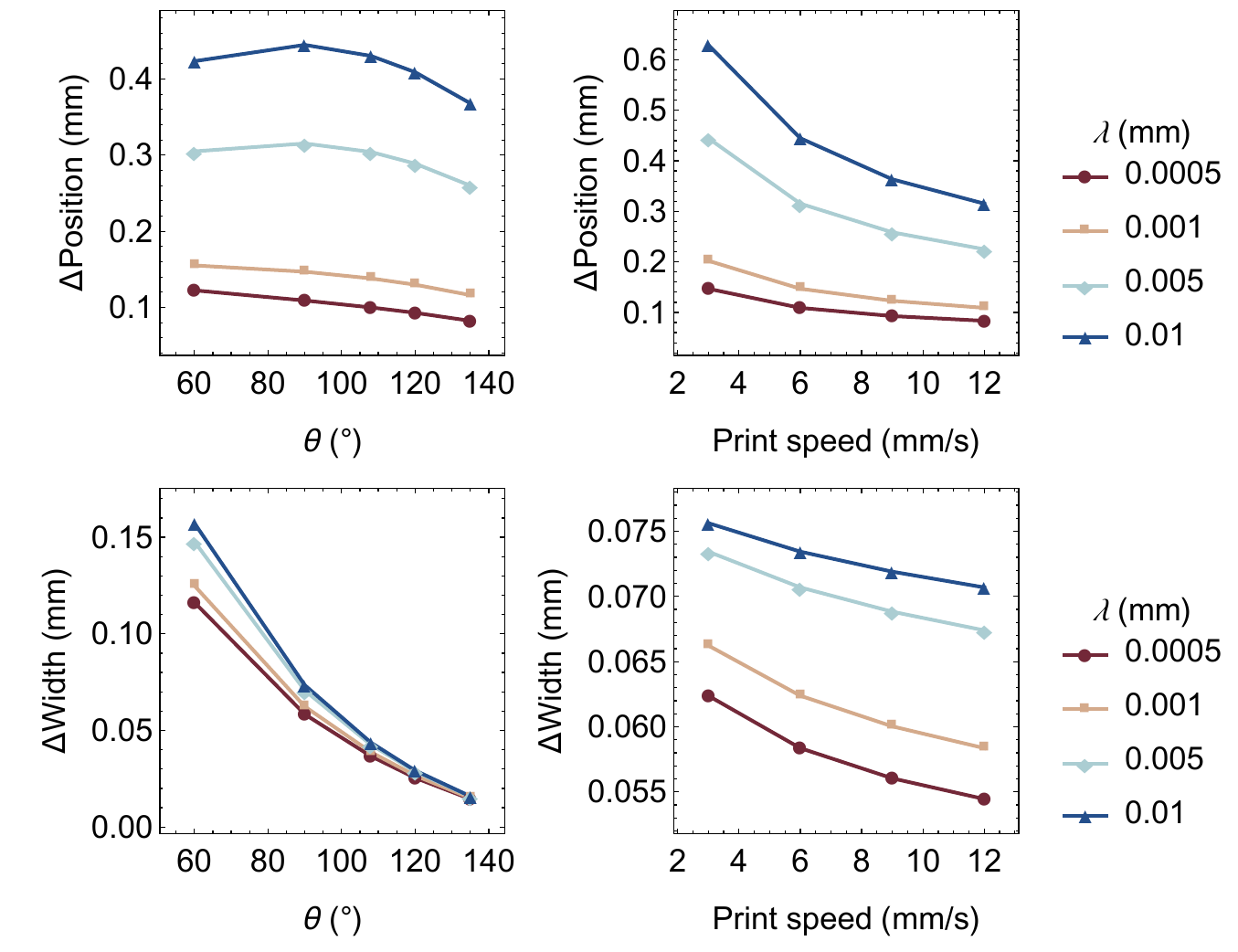}
 	\caption[Predicted defects as a function of spreading length $\lambda$]{Scaling of change in particle distribution position and width at the corner after shear due to capillarity for varying spreading length scales $\lambda$.}
 	\label{fig:lambdascale}
 \end{figure}
 
 \section{Corner swelling with acceleration}
 
 \begin{figure}[H]
 	\centering
 	\includegraphics{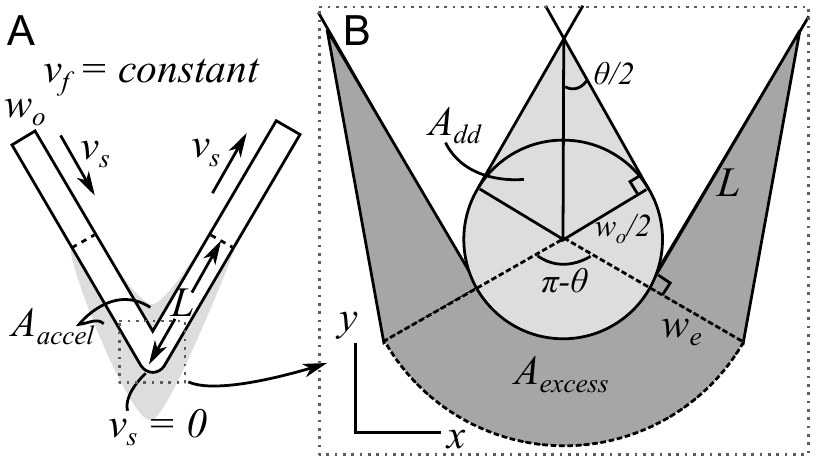}
 	\caption[Schematic of corner swelling]{Schematic of corner swell due to A) acceleration and B) double deposition.}
 	\label{fig:swellschem}
 \end{figure}
 
 There are two ways to structure the print path at a corner. First, the motor can cheat corners, sweeping an arc instead of a point. This is known as the \textit{blended-move} scheme, which results in corner smoothing.\cite{Comminal2019, Han2002} Second, the motor can slow to a speed $v_c$ (the jerk speed or ramp speed) at the corner, then accelerate. The jerk speed $v_c$ depends on hardware capabilities and software settings, since the 3-axis gantry can only achieve a certain instantaneous acceleration, and jerk speeds which are too high will cause ringing. The Shopbot Desktop 3-axis stage used in these experiments uses this second method, known as \textit{stop-at-turn}, which results in corner swelling.\cite{Comminal2019, Han2002} \textit{Stop-at-turn} produces accurate nozzle positioning, but because the nozzle continues to extrude fluid at the same rate while the translation speed decreases, extra fluid is extruded at the corner (Fig. \ref{fig:swellschem}A). The model used in the main body of the text assumes that the jerk speed is the same as the print speed, which is true for the experiments covered in this paper but will often not be true in other printing scenarios. Here, an analytical model which includes acceleration at the corner is described.
 
 The amount of time that the nozzle spends decelerating to a translation speed of $v_c$ and accelerating at the corner $t$ is
 
 \begin{equation}
 t_{accel} = \frac{2(v_s-v_c)}{a}
 \end{equation}
 
 Within that deceleration/acceleration zone, the average translation speed is $(v_s-v_c)/2$. As such, the length of the deceleration zone, which is equal to the length of the acceleration zone $L$, is
 
 \begin{equation}
 L = \frac{v_s-v_c}{2}\frac{t_{accel}}{2} = \frac{(v_s-v_c)^2}{2a}
 \end{equation}
 
 Under ideal conditions where the nozzle can instantaneously change speed, the amount of time spent in the zone of the same length
 
 \begin{equation}
 t_{ideal} = \frac{2L}{v_s} = \frac{(v_s-v_c)^2}{v_s a}
 \end{equation}
 
 As such, the added time spent in the acceleration/deceleration zone is 
 
 \begin{equation}
 \Delta t_{excess} = t_{accel} - t_{ideal} = \frac{v_s^2 + v_c^2}{v_sa}
 \end{equation}
 
 During that added time, the nozzle extrudes an excess volume $V_{accel}$ of ink
 
 \begin{equation}
 V_{accel} = v_f A_{nozzle} \Delta t_{excess}
 \end{equation}
 
 The experiments in this work use a flow speed equal to the translation speed, $v_f = v_s$ and a square nozzle of inner width $w_o$. As such,
 
 \begin{equation}
 V_{accel} = \frac{(v_s^2 - v_c^2)w_o^2}{a}
 \end{equation}
 
 In addition to acceleration, double deposition produces excess fluid (Fig. \ref{fig:swellschem}B). Double deposition occurs because every time the print path changes direction, the nozzle retraces some area that it already covered. Conventionally, the double deposition area is calculated assuming that the nozzle is circular with radius $w_o/2$ (Fig. \ref{fig:swellschem}B). If the nozzle stops moving at the corner, the resultant traced path will have an outer radius of $w_o/2$. In the experiments in this paper, the nozzle has a square cross-section, so the corner should match the square cross-section of the nozzle. However, the orientation of that square would vary based on the orientation of the corner. On average, we assume that the shape of the corner is a circle with radius $w_o/2$. 
 
 The double deposition area $A_{dd}$ can be expressed in terms of the corner angle $\theta$ and the corner width $w_o$.\cite{Han2002}
 
 \begin{equation}
 A_{dd} = 2(1/2)(w_o/2)(w_o/2\cot(\theta/2)) + (\pi+\theta)\pi(w_o/2)^2
 \end{equation}
 
 Assuming that the height of the deposited line is the stand-off distance $h$, the excess volume from double deposition $V_{dd}$ is thus
 
 \begin{equation}
 V_{dd} = \frac{w_o^2h}{4}\Big(\cot\Big(\frac{\theta}{2}\Big) + \frac{\pi}{2}+\frac{\theta}{2}\Big)
 \end{equation}
 
 The total excess volume of fluid is 
 
 \begin{equation}
 V_{excess} = V_{dd} + V_{accel}
 \label{eq:vtots}
 \end{equation}
 
 Ideally, the excess fluid would be deposited on the outer edge, in a sharp tip that falls within the intended print path.\cite{Han2002} However, the possibility of this sharp tip disagrees with numerical models and experimental results that indicate that the printed corner exhibits a rounded tip.\cite{Comminal2019, Kulkarni1999, Kao1998} Numerical models indicate that some of this excess volume will fall inside of the intended print path, and some will fall outside.\cite{Comminal2019} For a 90$^\circ$ corner, the volume of excess fluid deposited outside the print path corner is 2--3 times the volume deposited inside the corner.\cite{Comminal2019} This ratio expands to 10 times for 30$^\circ$ corners.\cite{Comminal2019} There is no accurate analytical model which predicts the ratio of excess volume deposited inside the corner to outside the corner. For simplicity, we will assume that all of the fluid is deposited outside of the corner in an arc of thickness $w_e$ that borders the circular shape of the corner, flanked by two triangles of height $L$ and width $w_e$, where $L$ is the length of the deceleration/acceleration zone. This entire area has height $h$, which is the stand-off distance between the nozzle and the substrate. Using Equation \ref{eq:vtots}, 
 
 \begin{equation}
 \frac{V_{excess}}{h} = \frac{\pi-\theta}{2\pi}\pi\Big(\Big(w_e+\frac{w_o}{2}\Big)^2-\Big(\frac{w_o}{2}\Big)^2\Big)+w_eL
 \end{equation}
 
 Solving for the arc width $w_e$,
 
 \begin{equation}
 w_e = \frac{-((\pi-\theta)w_o/2+L) +\sqrt{((\pi-\theta)w_o/2+L)^2 + 2(\pi-\theta)V_{excess}/h}}{\pi-\theta}
 \end{equation}
 
 The difference in line width between the corner and the center of the edge is $w_e$. The difference in the position of the center of the line between the corner and center is $-w_e/2$, because the center of the line shifts outward toward negative positions. 
 
 As the corner angle increases, the excess volume from double deposition decreases. However, the arc length $\pi-\theta$ decreases while $L$ remains the same, so more volume must be fit into a smaller length along the outside of the corner. As a result, $w_e$ increases with corner angle. Note that this is the opposite of the trend predicted by double deposition alone and does not describe the experimental data collected in this paper.
 
 \section{Rheology}
 
 \begin{figure}[H]
 	\centering
 	\includegraphics{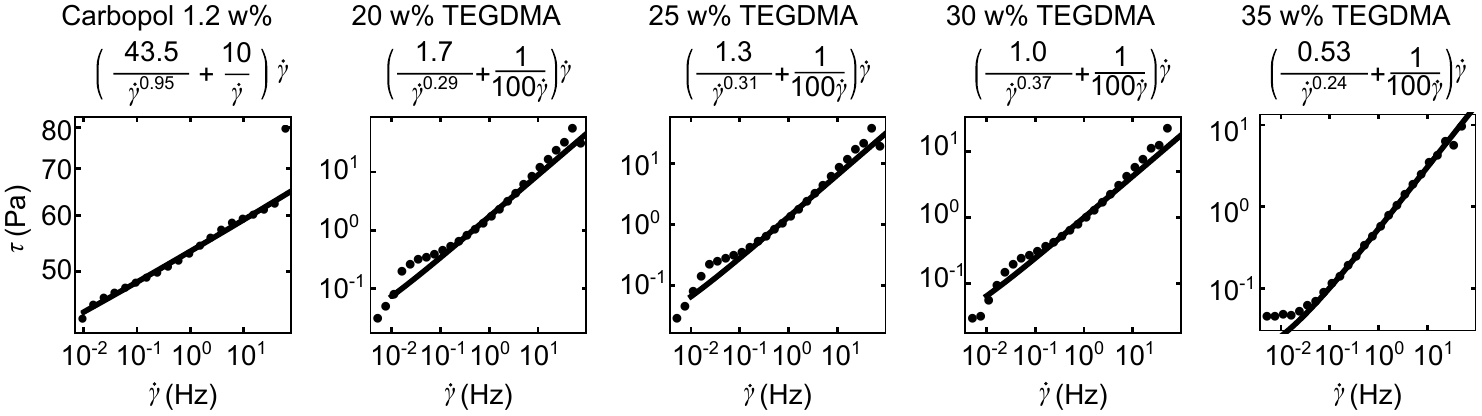}
 	\caption[Rheology of inks and support]{Viscosity as a function of frequency for Carbopol support and inks with varying w\% TEGDMA, fitted to the Herschel Bulkley model.}
 	\label{fig:rheology}
 \end{figure}
 
 Viscosities were measured using a TA Instrument Company ARES-LS1 rheometer with 25 mm diameter flat plates and a 2 mm gap at room temperature. Dynamic frequency sweeps were conducted at 10\% strain with increasing frequency. Dynamic strain sweeps were conducted at 10 Hz with increasing strain. To avoid curing and settling effects, inks did not contain photoinitiators or particles.
 
 The inks used in this study are liquid-like at all stresses, whereby the loss modulus is always greater than the storage modulus (Fig. \ref{fig:rheology}B). These inks are shear thinning but have a weaker rate dependence than the support material. In these inks, TEGDMA acts as a diluent, so increasing the TEGDMA concentration decreases the loss modulus and viscosity (Fig. \ref{fig:rheology}). 
 
 \section{Ink composition effects on changes at the corner}
 
 \begin{figure}[H]
 	\centering
 	\includegraphics{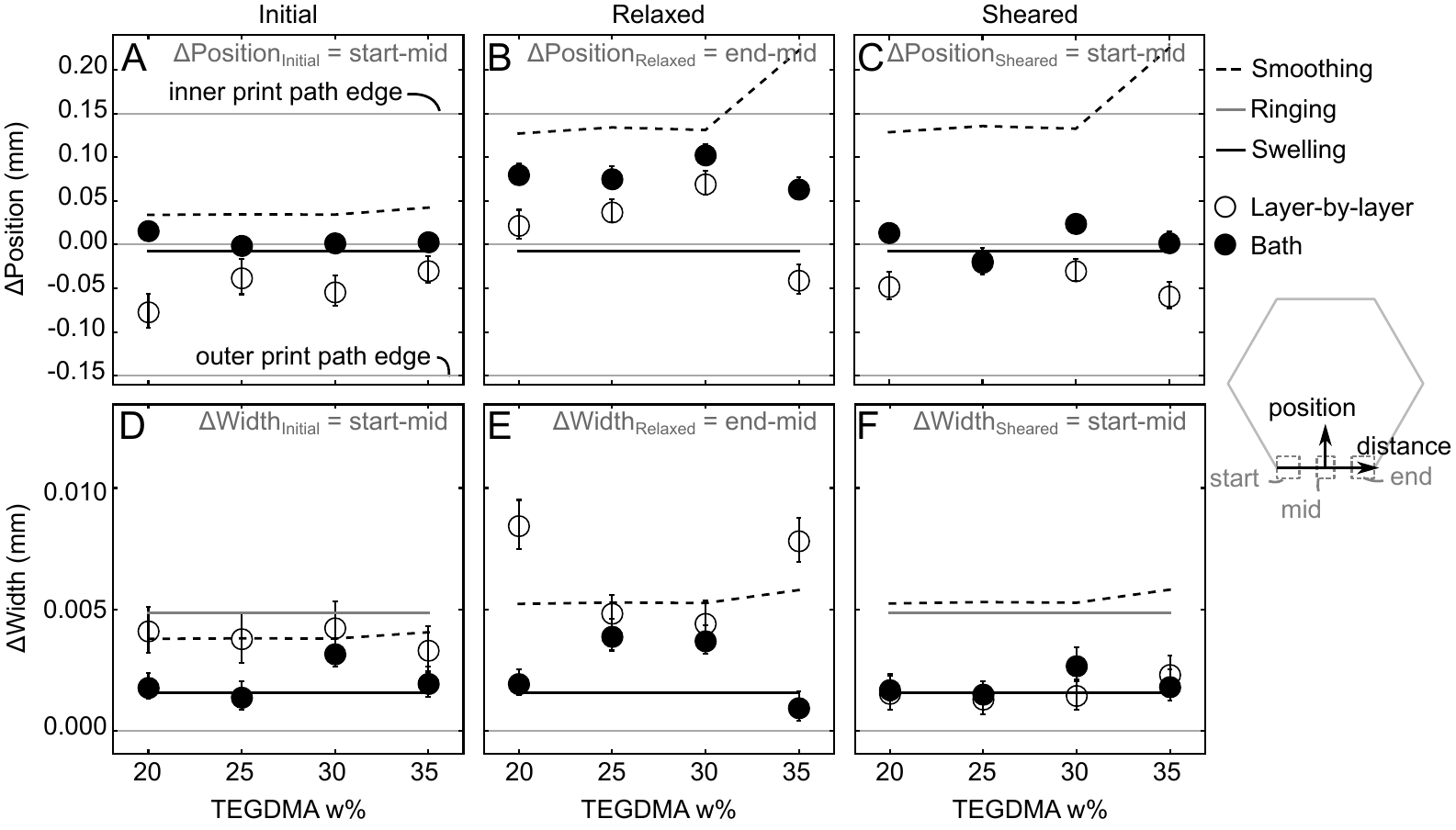}
 	\caption[Corner defects as a function of ink composition]{Change in position and width of the particle distribution at the corner as a function of ink composition. Theoretical contributions of the smoothing, swelling, and ringing models are shown. Error bars indicate standard error.}
 	\label{fig:timetheoryteg}
 \end{figure}
 
 The smoothing model only predicts a subtle dependency of change in width and position at the corner as a function of ink composition. At the highest TEGDMA loading and thus lowest viscosity, the smoothing model predicts an inward, positive shift in change in position and an increase in change in width. Experimental data indicate no strong dependence of change in position and change in width on TEGDMA content.

\end{document}